 \documentclass[preprint2,letterpaper]{aastex}

\slugcomment{To appear in ApJ, Draft version: \today}

\shorttitle{Surface Densities Correlation}
\shortauthors{Gutermuth et al.}

\begin{document}

\title{A Correlation Between Surface Densities of Young Stellar Objects and Gas in Eight Nearby Molecular Clouds}

\author{R. A. Gutermuth\altaffilmark{1,2}, J. L. Pipher\altaffilmark{3}, S. T. Megeath\altaffilmark{4}, P. C. Myers\altaffilmark{5}, L. E. Allen\altaffilmark{6}, \& T. S. Allen\altaffilmark{4}}


\altaffiltext{1}{Five College Astronomy Dept., Smith College, Northampton, MA 01063}
\altaffiltext{2}{Dept. of Astronomy, University of Massachusetts, Amherst, MA}
\altaffiltext{3}{Dept. of Physics \& Astronomy, University of Rochester, Rochester, NY}
\altaffiltext{4}{Dept. of Physics \& Astronomy, University of Toledo, Toledo, OH}
\altaffiltext{5}{Harvard-Smithsonian Center for Astrophysics, Cambridge, MA}
\altaffiltext{6}{National Optical Astronomy Observatories, Tucson, AZ}

\begin{abstract}
We report the discovery and characterization of a power law correlation between the local surface densities of Spitzer-identified, dusty young stellar objects and the column density of gas (as traced by near-IR extinction) in eight molecular clouds within 1~kpc and with 100 or more known YSOs.  This correlation, which appears in data smoothed over size scales of $\sim$1 pc,  varies in quality from cloud to cloud; those clouds with tight correlations, MonR2 and Ophiuchus, are fit with power laws of slope 2.67 and 1.87, respectively.  
The spread in the correlation is attributed  primarily to local gas disruption by stars that formed there or to the presence of very young sub-regions at the onset of star formation.
We explore the ratio of the number of Class~II to Class~I sources, a proxy for the star formation age of  a region, as a function of gas column density; this analysis reveals a declining Class~II to Class~I ratio with increasing column density.
We show that the observed star-gas correlation is consistent with a star formation law where the star formation rate per area varies with the gas column density squared. 
We also propose a simple picture of thermal fragmentation of dense gas in an isothermal, self-gravitating layer as an explanation for the power law. 
Finally, we briefly compare the star gas correlation and its implied star formation law with other recent proposed of star formation laws at similar and larger size scales from nearby star forming regions.


\end{abstract}


\section{Introduction\label{intro}}

It has been long recognized that the density of young stars varies by orders of
magnitude between different star forming regions.  \citet{herb62} compared the
low density of young stars in the Taurus star forming region to the high
density of young stars in the Orion nebula and speculated that these may form
an evolutionary sequence, with the low density regions collapsing into denser
structures.   More recently, ground-based near-IR surveys of molecular clouds
\citep[e.g.][]{lada91,stro93,carp00,porr03,lada03} identified embedded dense
young stellar clusters and groups in molecular clouds as well as a more diffuse
population of low mass stars, showing that molecular clouds contain both a
dense ``clustered'' and a diffuse ``distributed'' population.  Although
clusters could easily be identified in these surveys by searching for peaks in
the surface density of sources, measurements of the number of distributed young
stars in the clouds were dominated by uncertainties in the amount of
contamination by field stars and the treatment of variable extinction through
each cloud \citep{carp00}.  This led to a number of questions regarding the
number and origin of young low mass stars outside of clusters \citep{lada91,
li1997,carp00}.   What fraction of the cloud population is found outside of
clusters? Are they born in clusters and dynamically ejected to their observed
positions \citep{evan09}?  Alternatively, if they are formed near their current
positions, why do molecular clouds exhibit large variations in the density of
young stars?  Are the variations in density related to variations in the Jeans
length within clouds \citep{teix06,gute09}?  Better observational
characterization of the relationship between the spatial distribution of YSOs
and their natal cloud material is needed to understand the nature of the
distributed component, constrain models of star formation in molecular clouds,
and ascertain the underlying physics that determine the density and rate of
star formation \citep[e.g.][] {bate05, krum07, myer09}.

Sensitive {\it Spitzer} mid-IR and 24~$\mu$m surveys of nearby young clusters
embedded in molecular clouds have enabled complete characterization of the full
range of stellar densities in a range of star foming environments
\citep[e.g.][]{alle07,evan09}.  Young stellar objects (YSOs) in the clouds are
identified by the presence of IR-excess emission from circumstellar disks
and/or envelopes.  Several groups have developed methods to use {\it
Spitzer}-determined YSO spectral energy distributions to characterize the
sources as protostars (Class~0, I, or flat-spectrum) or pre-main sequence stars with disks (Class~II or transition disks) \citep[e.g.][] {gute09,
evan09, rebu10}.  By employing strategies to minimize the contamination by
background galaxies with similar infrared colors \citep[e.g.][]{harv06, gute08,
gute09, evan09}, the {\it Spitzer} surveys have yielded reliable, high spatial
dynamic range distributions of YSOs in their natal clouds.  These observations
show that the clusters are peaks of a more extended population of young stars
which tend to follow the morphology of the filamentary molecular clouds
\citep{alle07,gute09,evan09}.  Thus, low density star forming regions are not
limited to Taurus-like regions, but are also found in molecular clouds forming
large clusters and high mass stars \citep{alle07}.

The wide range in observed YSO surface densities provides an opportunity to
study how the density of young stars varies with the observed star formation
efficiency and the properties of the co-spatial molecular gas.  Recent work
indicates that the star formation efficiency increases with the stellar
density.  \citet{evan09} showed that higher surface density YSO clusterings
tend to exhibit higher star formation efficiency (30\%) than lower density
surroundings (3-6\%).  In a {\it Spitzer} survey of YSOs in the W5 molecular
cloud, where a recent generation of star formation is proceeding on the
peripheries of the evacuated cavities of adjacent HII regions, \citet{koen08}
found star-formation efficiencies of $>$10-17\% for high surface density
clusterings and 3\% for lower density distributed regions, respectively.  The
fraction of {\it Spitzer} identified YSOs outside of clusters depends strongly
on the adopted functional definition of stellar cluster, as there is no strong
distinction in source spacings evident in the nearby YSO spacing data to
suggest distinct ``clustered'' and ``distributed'' star-forming modes
\citep{bres10}.  The continuous nature of the overall YSO surface density
distribution in nearby clouds motivates a different approach whereby the local
surface density of YSOs and the local column density of gas are compared.  This
approach does not require a functional definition of clustered or distributed
star formation.

A similar approach has been used with success in observations of external
galaxies.  \citet{kenn98} outlines previously published results concerning the
validity of the Schmidt Law of star formation in external galaxies, namely
$\Sigma_{SFR} = A \Sigma_{gas}^N$, where the $\Sigma$s refer to the observed
surface densities of star formation rate and gas respectively \citep{schm59}.
For a limited range of densities and galaxy types these studies found $N$ to be
between 1 and 2.  \citet{kenn98}, using H$\alpha$ emission as the star
formation tracer, and CO and HI as the gas tracers for spiral galaxies, and
Br$\gamma$ emission and FIR dust emission as star and gas tracers respectively
for infrared-selected starburst galaxies, show $N = 1.45 \pm 0.15$.  Note that
these observations measure the integrated star formation (as traced by recently
formed OB stars) across multiple molecular clouds, hence the extragalactic laws
should not be expected to apply directly to studies of low mass stars within
individual molecular clouds in our galaxy.  That stated, \citet{evan09} compare
$\Sigma_{SFR}$ against $\Sigma_{gas}$ for the seven nearby clouds surveyed by
the {\it Spitzer} c2d Legacy survey against the Kennicutt-Schmidt Law, and find
$\Sigma_{SFR}$ to be $\sim$20 times larger than predicted by the law for the
measured $\Sigma_{gas}$.

In this contribution, we perform a direct comparison between locally determined
mass surface densities of YSOs and gas.  
This technique reveals a power law correlation between the YSO surface density and the gas surface density in all clouds, which we refer to hereafter as the star-gas correlation.   Furthermore, using the ratio of protostars to pre-main sequence stars as an evolutionary indicatory, deviations from that correlation can be interpreted as regions of more or less time evolution in the star-gas correlation. 
We interpret these by adopting a star formation law of power law form and
calculating the integral of the star formation rate over time vs. the remaining
gas density.  We present a simple analysis which shows that the higher density
regions of a cloud produce stars which consume their local gas more rapidly,
and the relation between log $\Sigma_*$ vs. log $\Sigma_{gas}$ increasingly
deviates from a power law with time, particularly at high gas densities.
Nevertheless, the observed correlation can be reasonably well explained by an
underlying $\Sigma_{SFR} = A \Sigma_{gas}^{2}$ star formation law.
We examine the observed deviations from the correlation and argue that they
result from a combination of gas dispersion in rich clusters and non-coevality
within the molecular clouds.

In Section~\ref{obs}, we summarize the {\it Spitzer} YSO lists compiled from
prior observations, as well as the near-IR extinction maps (as a tracer for the
molecular gas surface density).  In Section~\ref{corr}, we demonstrate and
characterize the observed correlations.  In Section~\ref{discuss}, we present
some analysis of the correlation and its implications including: a simple
evolutionary model to demonstrate the robustness of an underlying star
formation law of power law form under the effects of early gas depletion; a
simple thermal fragmentation model that predicts the central values of the
ranges of power law index and normalizations observed; a brief comparison to
the Schmidt-Kennicutt star formation law for galaxies.  Finally, we summarize
our findings in Section~\ref{summary}.

\section{Data and Methodology\label{obs}}

Numerous surveys of nearby star formation regions have been performed using
{\it Spitzer} \citep[e.g.][]{gute04,mege04,jorg06,harv07,alle07}.  One of the
most useful products of these observations is a largely flux-limited census of
high probability young stellar object candidates as indicated by their excess
mid-infrared emission relative to a photospheric spectral energy distribution.
{\it Spitzer} data cannot be used to uniquely distinguish diskless
pre-main-sequence stars from field stars, but this limitation to the total YSO
census is relatively minor for young, embedded star forming regions where few
members have had sufficient time for their disks to disperse
\citep[10-50\% for ages $<$1 to 3~Myr;][]{hern08}.  

Table~\ref{littab1} lists the cloud surveys we have used for this analysis, the
references for these surveys, and some general statistics for each.  We
selected large clouds with 100 YSO members or more, with a known range of YSO
surface densities that is greater than one order of magnitude, and located
within 1 kpc of the Sun, as these typically are complete for infrared-excess
sources down to roughly 0.2~$M_{\odot}$ at an age of 1~Myr \citep{barr98}, and
to even lower masses ($<0.08~M_{\odot}$) for the closer clouds.  However, the
presence of bright sources as well as bright, structured nebulosity near the
centers of some clusters does significantly reduce the completeness relative to regions of lower YSO density.  The most significant example of this issue is the Trapezium area
of Orion, where \citet{mege11} have performed a detailed comparison between the
{\it Spitzer} and {\it Chandra} source lists to examine the issue of
incompleteness.  
The completeness decreases rapidly in the inner regions of the Orion nebula: the comparison has revealed that the Spitzer survey may identify 50\% of the dusty YSOs averaged over the inner 0.5 parsec radius, and only identify 30\% of the YSOs in the inner 0.2 pc. 
Other expected trouble spots include
the centers of the MonR2 and GGD~12-15 clusters in the MonR2 cloud, and the
center of the Cep A cluster in the CepOB3 cloud.  In these regions of generally
high YSO surface density, we are likely underestimating the true surface
density of YSOs.   

The adopted datasets are relatively uniform in their observing parameters, but
heterogeneous in their data treatment, including different software pipelines
used for mosaic construction, source extraction algorithms, photometric
measurement techniques, and finally source classification schemes.  While the
results from the first three items are expected to be similar, it is not
intuitive to ascertain the degree of equivalence of the YSO type classification
schemes.  Some investigations into this issue have suggested that the relative
numbers of Class I and II YSOs are found to vary at the 10\% level by technique
\citep{alle07,rebu10,gute09}; method dependent variation at that level should
not strongly affect the results presented here.

\subsection{Large-scale Near-IR Extinction Mapping} 

To characterize the spatial distribution of gas column density in the target
molecular clouds, we have adopted dust column density maps from the near-IR
reddening of stars in the field of view, assuming that the dust and gas column
densities are proportional.  The maps are constructed following the method
described by \citet{gute09} for the Orion A and B, MonR2, CepOB3, S140, and
North America Nebula (NANeb hereafter) clouds.  For the c2d survey clouds
presented (Ophiuchus, Perseus, and Serpens), the extinction maps delivered as
part of that survey's enhanced data products have been adopted, after a
comparison between overlapping regions of the c2d maps and the maps of
\citet{gute09} showed consistent results (see Appendix~\ref{app1}).  The former
set of maps were generated at the Nyquist sampling resolution for the median
$n=20$ nearest neighbor distance in each region.  The sampling resolutions of
each map are tabulated in Table~\ref{littab1}, though it is worth noting that
the effective resolution of these maps decays somewhat at higher column
density.  The mean and maximum values after baseline subtraction, and the
value subtracted as a baseline, are also included in the table.  The empirical
extinction law of \citet{rl85} was adopted to convert the $H-K_S$ color
excesses to $A_V$ ($A_V/E\{H-K_S\} = 15.9$).  The gas column density maps are
computed from the extinction maps by assuming a constant gas to dust ratio of
$A_V/N(H_2)$ = 10$^{-21}$ magnitudes-cm$^2$.  This ratio has been observed to
vary by up to a factor of two \citep[e.g.][]{wins10}, however, such variation
will have a modest effect on the star-gas correlation which is observed over
more than an order of magnitude range in the gas column density.  

\subsection{YSO Surface Density Estimation}

We have performed a $n^{th}$ nearest neighbor surface density analysis of the
{\it Spitzer}-identified YSO distribution in each cloud, similar to the
analysis employed in \citet{gute09}.   For these values, $n$ was set to 11 -
this value results in uncertainties of 33\% in the surface density estimation
\citep{ch85}.   We first obtained these nearest neighbors centered on each of
the identified YSOs; for a sample of $q$ YSOs this would result in $q$ surface
density measurements centered on the YSOs.  We also obtained nearest neighbor
distances for a uniform rectangular grid of points in RA and Dec; this is the
same procedure used to generate nearest neighbor density maps in several recent
papers \citep[e.g.][]{gute05,roma08,chav08,jorg08,gute08,evan09,gute09}.  To
facilitate comparisons between the stellar density and the gas column density,
the grid vertices we have adopted here are identical to the extinction map's
pixel positions for each region.  The combined analysis yields two surveys of
the gas and stellar surface densities, one that follows the YSO spatial
distribution and the other that is unbiased and spatially regular.  In both
cases, the resulting densities of YSOs per solid angle in the sky are converted
to mass per square parsec using the distances in Table~\ref{littab1} and
adopting an average stellar mass of $0.5~M_{\odot}$.

\section{An Observed Correlation in the Surface Density of YSOs and Dust\label{corr}} 

\begin{figure}
\epsscale{.80}
\plotone{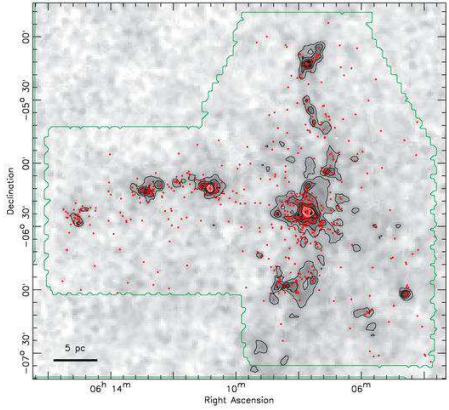}
\caption{Extinction map of the MonR2 cloud overlaid in red with the spatial distribution of {\it Spitzer}-identified YSOs.  The inverted grayscale is a linear stretch from $A_V~=~-1$ to 10 magnitudes.  Contour overlays start at $A_V=3$~mag and their interval is 2 mag.  The IRAC coverage is marked by the green boundary.  The projected positions of the YSOs in MonR2 closely trace almost all of the areas of detectably elevated extinction. Denser clusters of YSOs are clearly apparent in the zones of highest extinction.\label{monr2_im}}
\end{figure}

\begin{figure}
\epsscale{.80}
\plotone{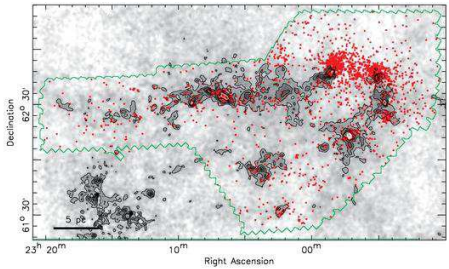}
\caption{Extinction map of the CepOB3 cloud overlaid in red with the spatial distribution of {\it Spitzer}-identified YSOs.  The grayscale and contour properties are identical to those in Fig.~\ref{monr2_im}.  As in that figure, YSOs are predominantly projected on the elevated extinction zones within the cloud, and clusters are found in the highest extinction zones.  However, unlike MonR2, the large CepOB3b young cluster in the northwest corner of the coverage is largely offset from significant extinction.  Focussed examination of this region in particular suggests that the OB stars present have dispersed much of the local natal cloud material \citep{getm09,alle11}.\label{cepob3_im}} 
\end{figure}

\begin{figure}
\epsscale{.80}
\plotone{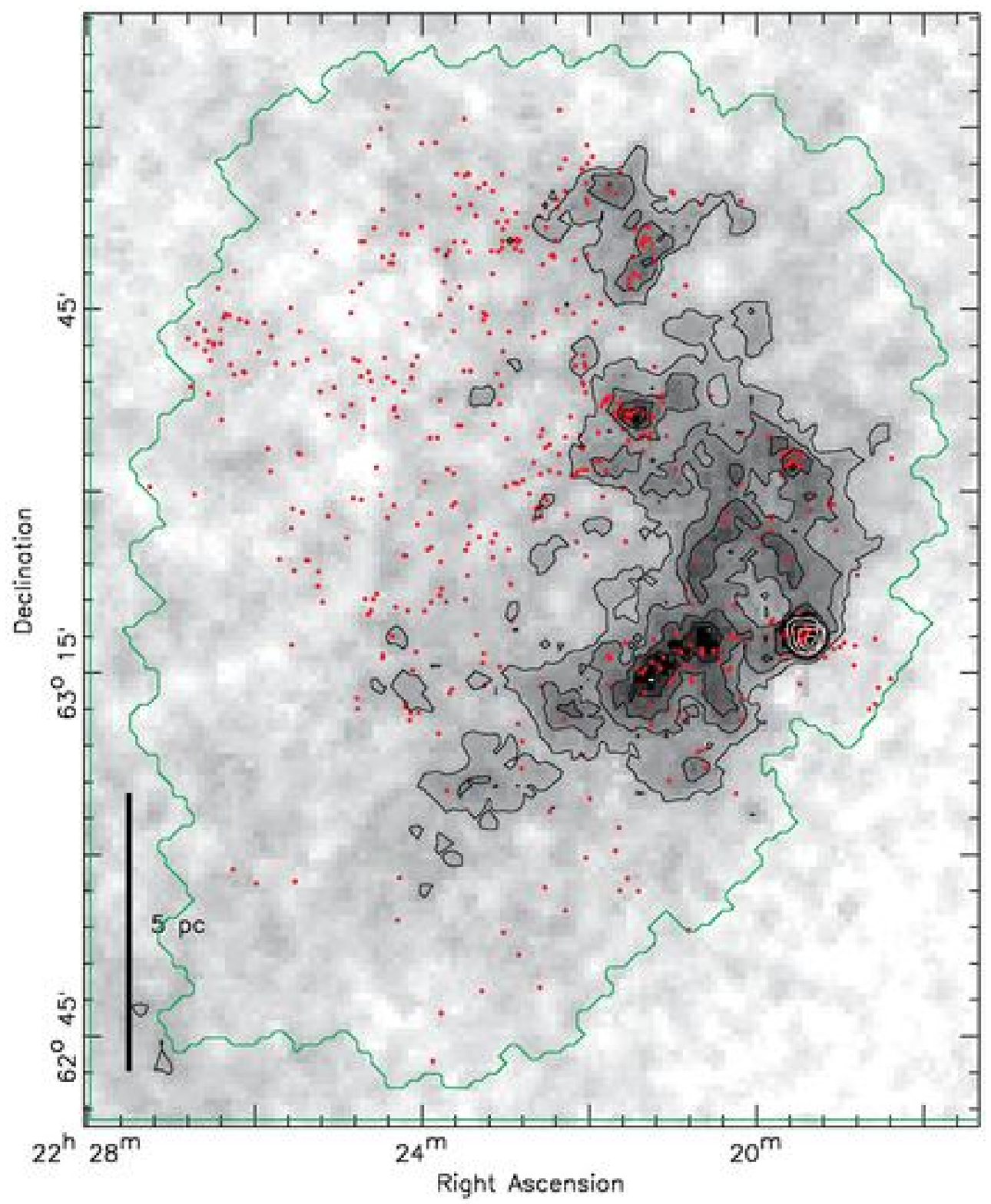}
\caption{Extinction map of the S140 cloud overlaid in red with the spatial distribution of {\it Spitzer}-identified YSOs.  The grayscale and contour properties are identical to those in Fig.~\ref{monr2_im}.\label{s140_im}}
\end{figure}

\begin{figure}
\epsscale{.50}
\plotone{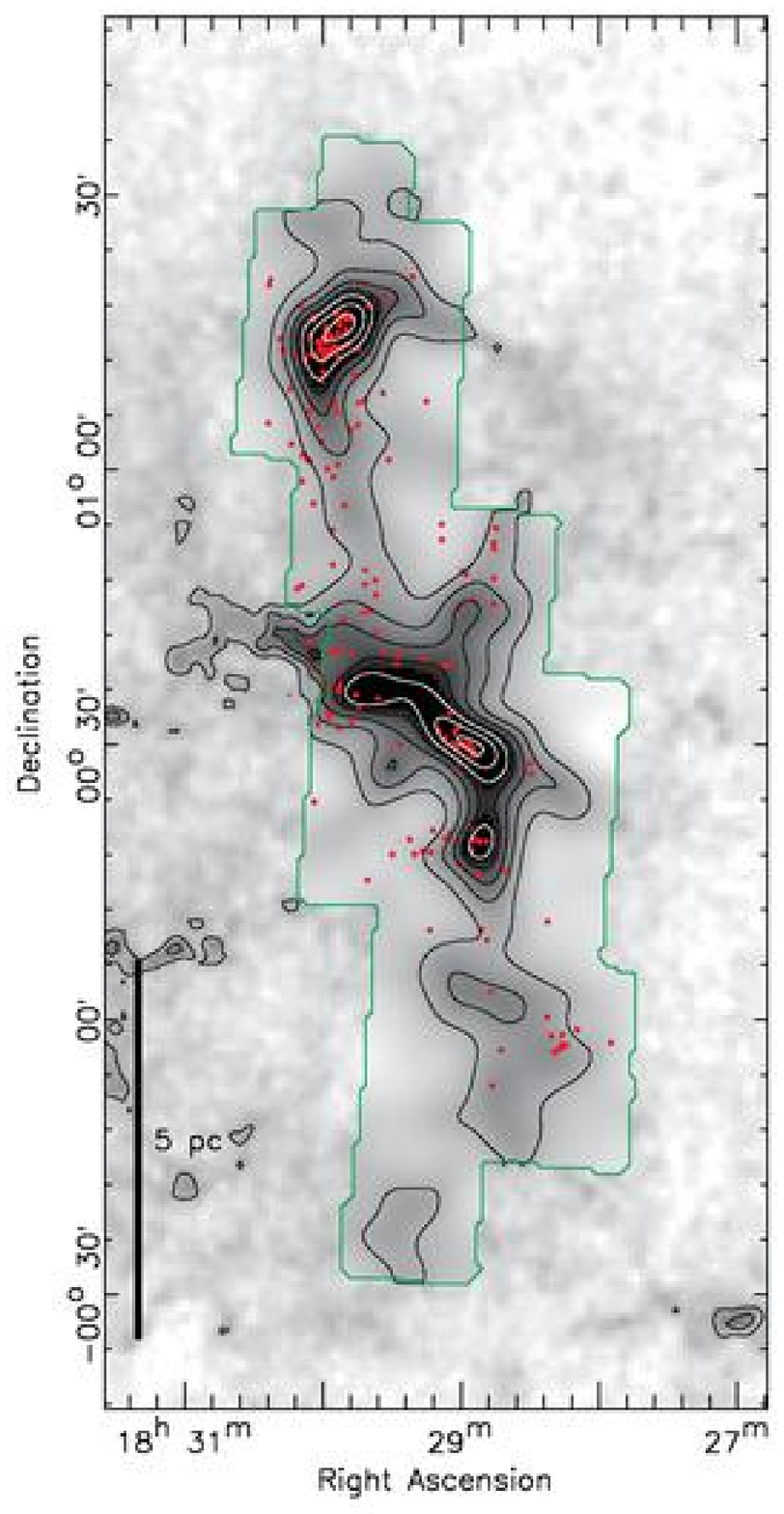}
\caption{Extinction map of the Serpens cloud overlaid in red with the spatial distribution of {\it Spitzer}-identified YSOs.  The grayscale and contour properties are identical to those in Fig.~\ref{monr2_im}.\label{serp_im}}
\end{figure}

Visual comparisons between the distribution of {\it Spitzer}-identified YSOs
relative to maps of gas structure in their natal molecular clouds have shown
that there is considerable similarity between them
\citep[e.g.][]{mege04,alle07,evan09,gute09}.  In Figure~\ref{monr2_im}, we show
one of the best examples of cloud-YSO correlation, the MonR2 cloud.  The red
dots are the positions of {\it Spitzer}-identified YSOs, the grayscale image is
the 2MASS-derived extinction map, and the green line outlines the {\it Spitzer}
mid-IR survey coverage.  The YSOs appear to congregate in the regions of
detectable extinction, peaking in density at or near the positions of highest
extinction.  Figure~\ref{cepob3_im} is the same figure, but for the CepOB3
cloud.  Here, the apparent correlation is similar if one considers only the
regions of significant extinction.  However, there is an exposed young cluster,
CepOB3b, in the northwest corner of the coverage where there are many YSOs
present and very little local extinction; the gas in this region has likely
been dispersed by winds and radiation from associated high mass stars.  (See Figures~\ref{s140_im}-\ref{serp_im} for similar figures for the rest of the clouds.) 

\begin{figure}
\epsscale{1.0}
\plotone{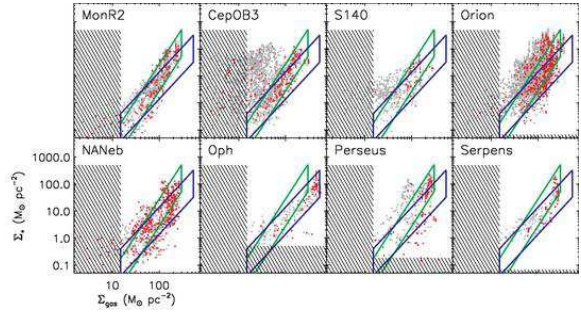}
\caption{YSO surface density versus gas column density for the eight molecular clouds considered here.  YSO surface density is computed from the local NN11 surface density of YSOs sampled at the position of each source.  Gas column density is derived from the nearest sample to each YSO's position in the extinction map.  Black hashed overlays mark out the regions of $A_V<1$ (ie. consistent with zero, given typical uncertainties) and where the angular surface density of sources is less than 7 per square degree (consistent with residual contaminant sources according to the \citet{gute09} classification scheme).  The green polygon marks the fit to, and estimated spread in, the MonR2 cloud's data, and the blue polygon is the same, but for the Ophiuchus cloud's data.\label{mfyso}}
\end{figure}

\begin{figure}
\plotone{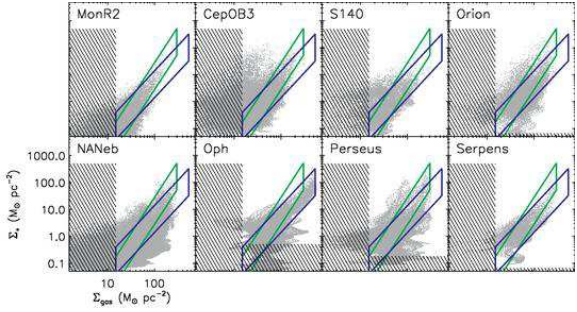}
\caption{Same as in Fig.~\ref{mfyso}, but now the stellar surface densities are sampled uniformly by position on the same uniform grid as the extinction map instead of centered on the sources themselves.  This area-sampled representation better represents regions of high gas column density but low YSO surface density.\label{mfarea}}
\end{figure}

To quantify the apparent correlation, we plot (on a log-log scale) the YSO mass
surface density $\Sigma_*$ vs. the molecular gas mass column density
$\Sigma_{gas}$ for MonR2, CepOB3, and six other star-forming clouds in
Figures~\ref{mfyso}~\&~\ref{mfarea}, utilizing the data and conversion factors
described above.  In Fig.~\ref{mfyso} for YSO-centered sampling, red points
represent central YSOs classified as protostars (hereafter Class~I, though
Class~0 and flat spectrum sources are included), and grey points represent
those that are classified as pre-main sequence stars with disks (hereafter
Class~II, though transition disks are included).  The black hatched regions
represent regions in the plot space where the extinction is too low to be
reliably measured in the near-IR at these resolutions (specifically $A_V<1$)
and where the surface density of stars is consistent with residual
contamination from AGN \citep[7 deg$^{-2}$;][]{gute09}\footnote{This value is
higher than the value reported for c2d \citep[e.g.][]{harv07}, but we apply it
across all clouds for consistency in treatment.}.  The bias toward sampling
only in positions with YSOs is addressed in Fig.~\ref{mfarea}.  Here, plot
points are sampled uniformly by position on the grid defined by the extinction
map sampling, yielding an area-sampled, rather than YSO-sampled, view of the
correlation.

Regardless of the view used, every cloud examined exhibits a positive power law
trend in the surface density of stars and column density of gas.  The MonR2 and
Ophiuchus clouds stand out as having particularly cohesive loci relating these
surface densities (Pearson coefficients of the log-log data are 0.87 and 0.83,
respectively), thus we use their correlation data to characterize the trend in
detail.  Using the same fitting method used by \citet{flah07}\footnote{The
\citet{flah07} algorithm is an IDL implementation of $\chi^2$ minimization of
the two dimensional perpendicular offsets from a line fit, thereby implicitly
incorporating uncertainty in both axis quantitites, following the analysis of
\citet{pres92}.}, we fit lines to the log~$\Sigma_*$ vs log~$\Sigma_{gas}$ data
for MonR2 and Ophiuchus, finding them well fit with power law indexes of $2.67
\pm 0.02$ and $1.87 \pm 0.03$, respectively.  Uncertainties in these values are
simply the intrinsic slope uncertainty from the fit; we estimate a systematic
uncertainty of $<0.04$ based on a comparative analysis of Ophiuchus using both
our extinction mapping technique, and that of c2d (see Appendix~\ref{app1}).
The fits are overlaid as polygons (green/steeper and blue/shallower for MonR2
and Ophiuchus, respectively) on the measurements for all clouds in
Figs.~\ref{mfyso}~\&~\ref{mfarea} to demonstrate that the correlations for the
other clouds are largely consistent with a power law of index $\sim$2.
However, most clouds have some larger spread in the observed correlation
relative to MonR2 and Ophiuchus ($\sim$1 dex spread), and this fact deserves
closer examination.  We have included fit results for all of the clouds in
Table~\ref{resulttab0}, for reference.

The more complex relationships seen in the other clouds appear related to
evolutionary differences in the star-gas correlation among different
structures within these clouds.  For example, the main cluster core of the
IC~348 region is well known to be a more evolved population of YSOs, with
evolved disks \citep[e.g.][]{muen07}, a low disk fraction
\citep[e.g.][]{lada06}, a high ratio (27) of Class~II to Class~I sources
\citep[e.g.][]{gute09}, and low mean extinction relative to other young
clusters \citep[e.g.][]{jorg08,gute09}.  All of these observations indicate an
evolved YSO population and thus significant time for the YSOs and the
intermediate mass central members to have dispersed the local gas via feedback
processes.  In the YSO vs. gas surface densities diagram for Perseus, the
subset of points that are significantly over-dense in stars relative to the gas
density predominantly belong to the IC~348 exposed cluster core.  Perseus also
has several regions that are under-dense in stars relative to gas.  These
regions are typically located in the well-studied dark clouds in Perseus, such
as B5 and L1448, where the YSOs in these objects are predominantly confirmed
Class~0 and Class~I objects \citep{jorg08,enoc08} with very few more evolved
Class~II sources.  This indicates relative youth when compared to more typical
embedded clusters like NGC~1333, which has a more standard ratio of Class~II to
Class~I sources \citep[3.7;][]{gute09}.  Ultimately, this study reveals that
the Perseus cloud is an outlier relative to other nearby molecular clouds
because of its heterogeneous composition of evolutionary states among its
star-forming sites.

The Cep~OB3 giant molecular cloud (GMC) appears to be a simpler (yet more
numerous in YSOs) analogue of Perseus in that it has a subset of YSOs that
deviate from the observed trend, namely the large, exposed Cep~OB3b cluster
consisting of two sub-clusters, Cep~OB3b-east and Cep~OB3b-west \citep{alle11}.
Cep~OB3b-east has an abnormally large current star-formation efficiency
\citep[80\%;][]{getm09}.  The cluster is found to be adjacent to two massive
molecular clumps indicating that most of the natal gas has been dispersed
\citep{getm09,alle11}.  Interestingly, the similarity of heterogeneous
evolutionary states in CepOB3 and Perseus may continue into the very young
regime as well; there are some regions in the Cep~OB3 molecular cloud which are
overdense in gas relative to YSOs hinted at in the area-sampled analysis shown
in Fig.~\ref{mfarea}.  

\begin{figure}
\plotone{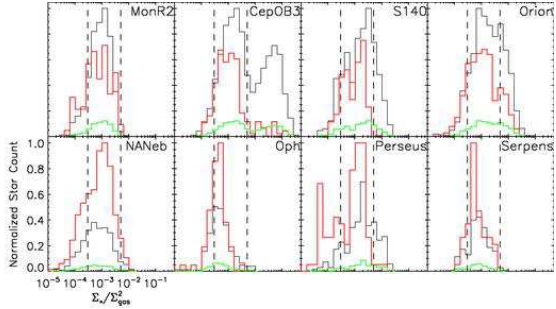}
\caption{Histograms of $\Sigma_*/\Sigma^2_{gas}$ using the data in Fig.~\ref{mfyso}, separated by YSO evolutionary class.  The Class~II YSO histogram is in black, the Class~I YSO histogram is in red, and each is normalized to their peak bin.  The green histogram is the Class~II histogram, but normalized to the Class~I histogram's peak bin and then divided by 30.  This represents the upper limit to the expected frequency that edge-on Class~II sources will be misclassified as Class~I YSOs \citep{gute09}.  Vertical dashed lines at $\Sigma_*/\Sigma^2_{gas} = 3 \times 10^{-4}$ and $5 \times 10^{-3}$ mark the empirical boundaries between different apparent evolutionary distinctions in the cloud: 
young, protostar-rich regions are on the left, most of the embedded cloud populations are found in the center and exposed, regions
that have undergone gas dispersal are on the right.
\label{histyso}}
\end{figure}

Given these qualitative regional evolutionary differences, we investigate the
degree that YSO evolution may trace differences in the
$\Sigma_*$~vs.~$\Sigma_{gas}$ analysis.  We visualize the differing Class~II to
Class~I ratios ($N_{II} / N_I$) of regions above or below the power law trend
by assuming that the power law index is always 2 and computing the implied
offset coefficient for the power law for each YSO individually.  These values,
$\Sigma_*/\Sigma^2_{gas}$, then may be composed into histograms separated by
YSO evolutionary class.  We have produced such histograms for all of the clouds
considered here (see Figure~\ref{histyso}).  The black histograms are the
Class~II YSO values, and the red histograms are the same, but for the Class~I
YSOs scaled up by 3.7, the median ratio of Class~II to Class~I sources in
nearby clusters \citep{gute09}.  Both have been normalized to the peak value
among both histograms.  The green histogram plots the Class~II YSO data, but
now scaled to the Class~I data's normalization and divided by 30, as $<1/30$
Class~II objects may be misidentified as Class~I YSOs due to an edge-on disk
orientation \citep{gute09}.  Internally coeval clouds like MonR2 and Ophiuchus
appear to have very similar distributions for both YSO classes.  In contrast,
those that are more heterogeneous in their evolution have a clear systematic
trend whereby there is an excess in the Class~I distribution where the
calibration value is low (gas-rich or YSO-poor) and an excess in the Class~II
distribution where the calibration value is high (gas-poor or YSO-rich).  

Empirically, we define $\Sigma_*/\Sigma^2_{gas} = 3 \times 10^{-4}$ and $5
\times 10^{-3}$~pc$^2$~M$^{-1}_{\odot}$ as boundaries that bracket the distributions of those clouds
with clear correlations and set apart the extrema of local evolution that
appear commonly in the more heterogeneous clouds.  Note that we have reduced
over three orders of magnitude variation in $\Sigma_*$ to an approximately one
order of magnitude variation in $\Sigma_*/\Sigma_{gas}^2$.
$\Sigma_*/\Sigma^2_{gas} < 3 \times 10^{-4}$ indicates a relative overdensity
in gas, and the stars that are present there are predominatly of the younger
type, Class~I.  We infer that the the onset of star formation was delayed in these regions, resulting in the presence of few YSOs and a large reservoir of molecular gas.  On the other extreme,
$\Sigma_*/\Sigma^2_{gas} > 5 \times 10^{-3}$ suggests stellar overdensity
relative to gas, and the stars found in this regime are nearly exclusively
Class~II YSOs.  Thus we infer that these are regions where star-forming gas has
been largely depleted by some combination of conversion into stellar mass or
ejection of gas via radiative processes, such as photoevaporation, or winds
from the recently formed stars.  We summariuze the median
$\Sigma_*/\Sigma^2_{gas}$ values by YSO class for each cloud in
Table~\ref{resulttab1}. 

\begin{figure}
\plotone{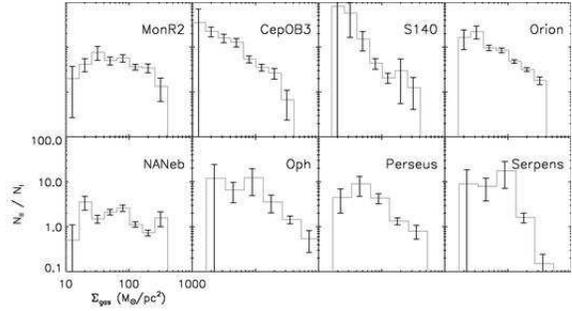}
\caption{Class~II to Class~I source count ratio versus gas column density for those objects with $3 \times 10^{-4} < \Sigma_*/\Sigma^2_{gas} < 5 \times 10^{-3}$. Among all clouds, there is a clear trend toward lower ratios at higher gas column densities.  This could suggest that low gas column YSOs are relatively more evolved if they formed in situ (which could be suggestive of either an earlier onset of star formation in the currently low column zones or that the YSOs in high column zones have longer Class~I phase lifetimes).  Alternatively, the low column YSOs may have migrated from high density regions or may be the result of gas dispersal.  This latter explanation may be particularly relevant for Cep OB3b, S140 and Perseus where there is clear evidence of gas dispersal.\label{c2c1vgas}}
\end{figure}

A majority of the YSO membership in each cloud (54\%-85\%) falls in the range
of $\Sigma_*/\Sigma^2_{gas}$ between $3 \times 10^{-4}$ and $5 \times
10^{-3}$~pc$^2$~M$^{-1}_{\odot}$.
This excludes regions of each cloud that have mostly dispersed their gas or
appear to be just at the onset of star formation.  Such a selection has been
applied to make Figure~\ref{c2c1vgas}, the Class II to Class I ratio of sources
as a function of gas column density for each of the clouds.  
This analysis yields a diagnostic of the evolutionary state of the molecular
clouds at different gas column densities.
Even within the restricted range of $\Sigma_*/\Sigma_{gas}^2$,  there are
significant variations in the ratio of  Class~II to Class~I objects.  Overall,
most clouds show a trend of decreasing Class II to Class I ratio with
increasing gas column density.  This may suggest that the higher column density
regions are younger than those with low column density.; however, there are
several other possible interpretations for this trend.  For example, the
protostar phase  may last longer in high density regions, or protostars in
these regions may be less prone to oscillate between Class II and Class I SEDs
\citep{dunh10}.  Alternatively, the regions of low column density may be
contaminated by Class~II objects that have migrated from clusters.  Gas
dispersal may result in the high fraction of Class II at low gas densities;
this may be especially important for the Cep OB3b, S140 and Perseus clouds.
Finally, we cannot rule out observational biases.  Although our adopted
classification criteria were designed to minimize the influence of extinction,
high extinction may result in the mis-classification of some Class~II YSOs as
Class~I protostars.  
%

\section{Discussion\label{discuss}}

The Class~II YSO evolutionary stage lasts between 0.5 and 5~Myr \citep{hern08},
and the majority of the IR-excess sources included in this analysis are of that
class.  Therefore, the fact that the above power law trend is observed at all
suggests a close link between the YSO spatial distribution and the gas
structure on the timescale of $\sim$2~Myr for the $\sim$1~pc size scales probed
here.  While this is indeed a profound result, it is still in some sense
unconstraining.  
The connection between the gas and the YSOs may in part result from the
observed kinematic connection between dense cores and the surrounding molecular
gas.  Specifically, millimeter-line observations show that the the velocity
dispersion between the  cores and the surrounding, moderate density gas is
small relative to the velocity dispersion inferred from the observed linewidth
and the observed velocity gradients in the bulk gas motions
\citep[e.g.][]{wals04,wals07,adam06,kirk07,jorg08}.  
However, the YSOs and their progenitor gas could simply co-move in aggregate
around the larger structure of the cloud, perhaps even dynamically interacting
with other such aggregations, and still preserve the observed connection
between the YSO and gas distribution.
Thus, a dynamical picture for the cloud and the structures within the cloud is
not ruled out by the star-gas correlation.

Under the simplifying assumption that there is relatively little motion between
the YSOs or the gas, we present a simple model which adopts a star formation
law where the star formation rate per area follows a power law of the local
instantaneous gas surface densities.  This model demonstrates that the likely
gas depletion that has occured to generate the observed star-gas correlation
still leaves behind a power law of similar index to the primordial
relationship.  
We then establish that the observed star-gas correlations and the star
formation law have a power law index $\sim$2. 
Next, we demonstrate that simple thermal fragmentation of an isothermal,
self-gravitating, modulated layer of gas can produce a spatial distribution of
gravitationally unstable fragments where the surface density of fragments
scales as the gas density squared.  Finally, we briefly compare the reported
star-gas correlation to similar recent analyses in the literature.

\subsection{Comparing Star Formation Laws to the $\Sigma_*$ vs. $\Sigma_{gas}$ Relationship\label{tommodel}}

As discussed in Section~\ref{intro}, several empirical star formation laws have
been proposed in the literature; these laws relate the inferred star formation
rate per area to the current column density of gas, and typically they are
derived over large ($>$100~pc) size scale measurements.  We examine here
whether the observed $\sim$1~pc scale $\Sigma_*$~vs.~$\Sigma_{gas}$
correlation reported in Section~\ref{corr} is consistent with a simple star
formation law where the star formation rate per area varies linearly or as a
power of the current gas column density.  

In order to consider the likelihood that the reported power law correlation in
observed quantities, ie. the integrated number of YSOs and the remaining amount
of gas, relates at all to an underlying star formation law of power law form,
we must consider whether the more efficient creation of stars at higher gas
column density significantly affects the underlying gas distribution and
therefore results in an observed $\Sigma_*$~vs.~$\Sigma_{gas}$ distribution
that deviates from the power law index of the star formation law itself.  To
this end, we present a simple model of gas to stellar mass conversion.  Gas
mass is accreted onto stars at a rate governed by an assumed star formation law
with an efficiency that accounts for the ejection of gas mass from the system
via protostellar outflows.  

The nature of this model demands several underlying assumptions to both achieve
adequate simplicity and offer a relatively robust examination of how the
observables behave with time.  The first assumption is that the stars formed do
not move significantly from their birth sites to within the typical resolution
of our measurements ($\sim$1~pc scale).  The observations that spatially
proximate prestellar cores are likely formed with low relative velocities
\citep{wals07,muen07b} suggest that the typical velocity dispersion of of the
YSOs that emerge from the cores is small ($\sim$0.4~km~s$^{-1}$).  Furthermore,
later in the star formation process, relatively uniform surface density
distributions are observed in many embedded YSO groupings \citep{gute09},
suggesting relatively little dynamical evolution.  The second assumption is
that there is no molecular gas flowing into or within the molecular cloud at
the $\sim$1~pc scale; gas is only ejected from the system by the formation of
stars, or converted into stellar mass.  This assumption is adopted for
simplicity; it is demonstrably false at scales $<$1~pc, where systematic gas
infall motions have been observed in cluster-forming clumps
\citep[e.g.][]{wm99,wals06}, and on larger scales, the global freefall time of
a typical GMC is $\sim$5~Myr, though turbulent support likely inhibits global
collapse to some degree.  Finally, we assume that there is no effect on the gas
distribution by large scale effects from phenomena such as feedback from nearby
high mass stars, supernovae, or galactic motions.  These processes only affect
the cloud on longer timescales ($>$10~Myr) than we are considering, with the
exception of high mass star feedback that we ignore for simplicity. 

We begin by adopting a general star formation law where the star formation rate
per area shows a power law dependence on the gas column density:

\begin{equation}
\frac{\partial \Sigma_{\star}(x,y,t)}{\partial t} = c k \Sigma_{gas}(x,y,t)^{\alpha}
\label{eqn:sflaw}
\end{equation}

\noindent where $\Sigma_{\star}(x,y,t)$ is the mass surface density of YSOs as
a function of position in the sky and time, $k$ is a constant that is defined
in equation~\ref{gasdepl} below, and $\Sigma_{gas}(x,y,t)$ is the mass column
density of molecular gas as a function of position and time.  We have assumed
that the formation of a star depletes the cloud by $M_{\star}/c$ where
$M_{\star}$ is the mass of the star created and $c$ is the mass conversion
efficiency which takes into account the mass of gas ejected from the cloud by
feedback during star formation ($M_{ejected} = M_{\star}/c-M_{\star}$).  Here
we are assuming that the fraction of the mass in the outflow ejected from the
cloud is constant throughout the cloud, but in fact the fraction of mass that
escapes the cloud's gravitational potential may vary as a function of local gas
column density and be affected further by the nearby gas configuration at
$>$1~pc scales.  Furthermore, it is worth noting that in this formulation we
are not considering individual stars, thus the entire analysis, including the
value of $c$, is averaged over the stellar initial mass function (IMF),
regardless of its form.  A pre-stellar core to star mass conversion efficiency
that is constant over the stellar mass range is consistent with recent
characterizations of the mass function of pre-stellar cores
\citep[e.g.][]{alve07,andr10}.  They find a value for this efficiency of $0.3
\pm 0.1$.  However, this does not speak to the amount of unaccreted matter that
is ejected from the cloud entirely; thus we consider 0.3 as a lower limit on
the value of $c$. 

If we assume that no additional gas is flowing into or within the cloud as the
stars form, then the rate of depletion of the gas mass is given by:

\begin{equation}
\frac{\partial \Sigma_{gas}(x,y,t)}{\partial t} = - k  \Sigma_{gas}(x,y,t)^{\alpha}
\label{gasdepl}
\end{equation}

\noindent In the case that $\alpha \ne 1$, the solution for
$\Sigma_{gas}(x,y,t)$, the gas density remaining in the cloud, is:

\begin{equation}
\Sigma_{gas}(x,y,t) = \Sigma_{gas}(x,y,0) (\frac{t}{t_0}+1)^{\beta}
\label{eqn:siggas}
\end{equation}

\noindent where, $\Sigma_{gas}(x,y,0)$ is the column density of gas at the
onset of star formation, $\beta$ is  a constant, and $t_0$ is the timescale for
the gas to be depleted by a factor of $2^\beta$.  Specifically:

\begin{equation}
\beta = \frac{1}{1-\alpha}
\end{equation}

\noindent and

\begin{equation}
t_0 = \frac{1}{ k (\alpha-1) } \Sigma_{gas}(x,y,0)^{1-\alpha}
\label{t0def}
\end{equation}

\noindent Note that $t_0$ depends on the initial column density of gas for
$\alpha > 1$; in this case regions with higher column densities will be
depleted more rapidly.  The amount of mass converted into stars is then the
mass of the depleted gas times the mass conversion efficiency, $c$:

\begin{equation}
\Sigma_*(x,y,t) = c [\Sigma_{gas}(x,y,0) - \Sigma_{gas}(x,y,t)]
\label{eqn:siggas_diff}
\end{equation}

\noindent Substituting equation Eqn.~\ref{eqn:siggas} into
Eqn~\ref{eqn:siggas_diff}, we obtain the relationship:
 
\begin{equation}
\Sigma_{\star}(x,y,t) = c\Sigma_{gas}(x,y,0)[1-(\frac{t}{t_0}+1)^{\beta}].
\label{fullsoln}
\end{equation}

In the case that $t << t_0$, we find that $\Sigma_{\star}(x,y,t)$ has the same power law dependence as the star formation rate:

\begin{equation}
\Sigma_{\star}(x,y,t) =  c \Sigma_{gas}(x,y,0)^{\alpha} k t
\label{simplesoln}
\end{equation}

\noindent 
In other words, the surface density of formed stars is proportional
to the surface density of initial gas to the $\alpha$ power.   
In this limit, the observed gas density can be approximated as
 $\Sigma_{gas}(x,y,t) = \Sigma_{gas}(x,y,0)$ and the following
approximation for the star formation rate can be used:

 \begin{equation}
\frac{\partial \Sigma_{\star}(x,y,t)}{\partial t} = \frac{\Sigma_{\star}(x,y,t)}{t} =  ck \Sigma_{gas}(x,y,t)^{\alpha}
\label{alphagen}
 \end{equation}

\noindent
In the case that the star formation rate is proportional to the gas density,
i.e. $\alpha = 1$ (even though our data cannot be reproduced by such a law, as
we next show), we have a different solution to equation~\ref{gasdepl} for the gas remaining, $\Sigma_{gas}(x,y,t)$:

\begin{equation}
\Sigma_{gas}(x,y,t) = \Sigma_{gas}(x,y,0)e^{-kt}
\end{equation}

\noindent
and for $\Sigma_{\star}(x,y,t)$ we find the solution:

\begin{equation}
\Sigma_{\star}(x,y,t) = c \Sigma_{gas}(x,y,0)[1-e^{-kt}]
\end{equation}

\noindent
which becomes, in the limit of $kt << 1$:

\begin{equation}
\Sigma_{\star}(x,y,t) = c \Sigma_{gas}(x,y,0) k t
\label{alpha1}
\end{equation}

\noindent Eqn.~\ref{alpha1} thus simplifies to Eqn.~\ref{simplesoln} for the $\alpha=1$ case, under a similar assumption of small $t$. 

At this point, we have argued that a power law observed in $\Sigma_*$~vs.~$\Sigma_{gas}$ is indicative of a star formation law of similar functional form.  Therefore, it is interesting to examine the $\alpha =2$ case, as suggested by the analysis from Section~\ref{corr}.  

Combining Eqn.~\ref{eqn:siggas}~and~\ref{t0def} under the assumption that $\alpha=2$ allows us to find a useful expression for the quantity $\Sigma_{\star}/\Sigma_{gas}^2$ computed in Section~\ref{corr}:

\begin{equation}
\frac{\Sigma_{\star}}{\Sigma_{gas}^2}(x,y,t) = ckt (1+\frac{t}{t_0})
\label{sigrat}
\end{equation}

\noindent Thus for $\alpha=2$, the histograms presented in Fig.~\ref{histyso}
can be thought of as histograms of normalized star counts for Class I and II
YSOs vs.  $ckt_{SF}(x,y) (1+\frac{t_{SF}}{t_0}(x,y))$ at the current mean age
of the YSOs at a given position, $t=t_{SF}$.  Using this result and the
observed evolutionary phase ratios of the YSO population,
$N_{II}/N_I$~vs.~$\Sigma_{gas}(x,y,t_{SF})$ (Fig.~\ref{c2c1vgas}), we can
constrain the constant $k$ and thus the gas depletion time $t_0(\Sigma_{gas})$.
Then we can directly compare our data to the gas depletion model.

Here we measure $t_{SF}(x,y)$ by computing the ratio of the surface density of
stars formed divided by the star formation rate surface density.  We assume
that the star formation rate surface density is constant in time and that it is
approximated by the ratio of the surface density of Class~I objects over the
lifetime of that evolutionary phase,\ $\Sigma_I/t_I$, where $t_I\sim$0.5~Myr \citep{evan09}.  Thus:

\begin{equation}
t_{SF}(x,y) = \Sigma_*(x,y,t_{SF}) \times t_I / \Sigma_I(x,y)
\end{equation}

\noindent
$\Sigma_*(x,y,t_{SF}) = \Sigma_I(x,y)+\Sigma_{II}(x,y,t_{SF})$ in this analysis, thus we can simplify to:

\begin{equation}
t_{SF}(x,y) = (\frac{N_{II}(x,y,t_{SF})}{N_{I}(x,y)}+1) \times t_I
\label{tsfdef}
\end{equation}

As shown in section~\ref{corr}, we have computed $N_{II}/N_{I}$ for each cloud
as a function of gas column density, noting that some clouds have a downward
trend, while others are fairly stable over the whole range of density values.
The more extreme variations occur in clouds with wider spreads in $\Sigma_* /
\Sigma_{gas}^2$, thus we assume that those clouds have some contamination of
lower gas column density regions by older YSOs (via their motion away from
their birth sites) or that a large fraction of the primordial gas in those
regions has been recently ejected by feedback from nearby, high mass stars.
Given these complications, we have chosen to characterize the trend in
$N_{II}/N_{I}$~vs.~$\Sigma_{gas}$ from the relatively uniform MonR2 data,
yielding:

\begin{equation}
 \frac{N_{II}}{N_{I}}(x,y,t_{SF}) = 25 \times \Sigma_{gas}(x,y,t_{SF})^{-0.4}
 \label{n2n1}
\end{equation}

\noindent The observed data are largely confined to the range $20 <
\Sigma_{gas}(x, y, t_{SF} ) < 300$~M$_{\odot}$~pc$^{-2}$; evaluating
eqn.~\ref{n2n1} at these fiducial points yields $N_{II}/N_{I} = 7.5$~and~3,
respectively, and
$t_{SF} = 4.3$ and 1.8~Myr, respectively.  We now use these results  to set
limits on the value of $k$, and thus $t_0$.  

Since $t_{SF}$ and $t_0$ depend on quite different powers of $\Sigma_{gas}$,
the case of $t_{SF}<<t_0$ likely occurs at low initial gas column densities
(this will be verified later), thus the higher value, $t_{SF}=4.3$~Myr, is
applicable.  Assuming $c=0.3$ (the lower limit), then $2.3 \times 10^{-4} < k <
3.9 \times 10^{-3}$~Myr$^{-1}$~pc$^2$~$M_{\odot}^{-1}$.  Therefore, at the
corresponding value of $\Sigma_{gas}$ (20~$M_{\odot}$~pc$^{-2}$), where the
depletion time is long and thus $\Sigma_{gas}(x,y,t) = \Sigma_{gas}(x,y,0)$, we
find that the depletion time is $13 > t_0 > 220$~Myr\footnote{Note that $t_0$
would increase (as k would decrease) linearly if $c$ is larger, and $c \le 1$
by its definition above.}.  This range is 4 to 50 times larger than
$t_{SF}=4.3$~Myr and thus consistent with the $t_{SF}<<t_0$ assumption.
Assuming that the spread in the MonR2 $\Sigma_*/\Sigma_{gas}^2$ values are
dominated by data uncertainty\footnote{Our stellar surface densities are $n=11$
nearest neighbor densities, which are inherently uncertain at the 33\% level
\citep{ch85}, and our gas surface densities are uncertain at 0.5-4~$A_V$, with
higher uncertainty at higher surface densities.  These uncertainty sources
result in $1\sigma$ uncertainties of $\sim$0.3~dex in $\Sigma_*/\Sigma_{gas}^2$
values, thus they are sufficient to account for the $\pm0.6$~dex range in those
values in MonR2, assuming that they are log-normally distributed and that our
adopted limiting values isolate the central $\pm2\sigma$ range.}, we adopt
the log-scaled mean value for the $\alpha=2$ case,
$k=9.5\times10^{-4}$~Myr$^{-1}$~pc$^2$~$M_{\odot}^{-1}$. 

With $k$ adopted, we now determine the gas column density where $t_0 = t_{SF}$,
as this is the regime where significant gas depletion is expected in the data
and thus where deviations from a power law star-gas correlation would be
detectable.  For $\alpha=2$, $t_0$ is the time at which $\Sigma_{gas}(x,y,t_0)
= 0.5~\Sigma_{gas}(x,y,0)$.  Therefore:

\begin{equation}
t_0 = \frac{1}{2k} \Sigma_{gas}(x,y,t_0)^{-1}
\end{equation}

\noindent Setting $t_0=t_{SF}$ and thus this equation equal to
Eqn.~\ref{tsfdef} and solving numerically, we find
$\Sigma_{gas}(x,y,t_{SF}=t_0) = 300 M_{\odot}$~pc$^{-2}$.  Thus gas depletion
only becomes noticeably relevant at gas column densities on the high end of
the observed range.  

In summary, the model closely follows the observed power law correlation
between star formation rate surface density and gas column density.  Only at
the highest observed column densities does the model predict significant
deviations from a power law: these may be undetected due to the scatter in the
observed correlation.

\begin{figure}
\epsscale{.80}
\plotone{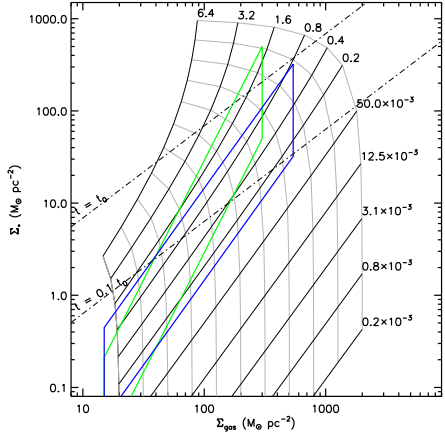}
\caption{Results from the simple gas depletion model for
an assumed star formation law of power law index 2 and an adopted star formation rate coefficient $k = 9.5 \times
10^{-4}$~Myr$^{-1}$~pc$^2$~$M_{\odot}^{-1}$.  Gray lines trace the star-forming evolution of parcels of gas of a given surface density through its evolution predicted by the model; as mass is converted from gas mass to stellar mass, the initial samples
drift upward.  For tracks that approach the regimes where $t \sim t_0$, gas is
significantly depleted by the star formation process and points arc to the left
as they rise.  The black lines are isochrones defining the locus of points for the coeval evolution of molecular gas spanning a range
of initial gas densities.
The ages
for each isochrone are noted, in Myr.  The colored polygons demarcate the area
that the observations of the MonR2 (green, steeper rise) and Ophiuchus (blue,
shallower rise) clouds occupy.  The dot-dashed lines mark where $t = t_0$ and
$t = 0.1~t_0$ for the model parameters chosen.\label{alpha2}} 
\end{figure}

At this point, it is useful to visualize the time evolution of the simple gas
depletion model and compare it with our observations.  In Fig.~\ref{alpha2}, we
plot the realizations of Equation~\ref{fullsoln} for several values of $t$ in
the range of 200 years to 6.3~Myr and a range of $\Sigma_{gas}(t=0)$ from 20 to
2000~$M_{\odot}$~pc$^{-2}$, assuming $\alpha$ = 2, $c$ = 0.5, and $k = 9.5
\times 10^{-4}$~Myr$^{-1}$~pc$^2~M_{\odot}^{-1}$.  These realizations are
isochrones showing the locus in the $\Sigma_*$ vs. $\Sigma_{gas}$ diagram for
an ensemble of co-eval molecular cloud parcels with different initial values of
$\Sigma_{gas}$.  We also plot orthogonal tracks that trace the evolution in
$\Sigma_*$ vs. $\Sigma_{gas}$ for parcels of a molecular cloud as the gas of a
given $\Sigma_{gas}(t=0)$ is converted into stars.  The polygonal regions that
demarcate the observed data limits for MonR2 (green) and Ophiuchus (blue) are
overplotted for comparison.  At early $t$, the model follows a power law of
index $\sim$2.  As we approach typical star formation region ages of
$\sim$1~Myr, the highest gas density evolutionary tracks curve toward
decreasing gas densities, resulting in local steepening of the model isochrone.
This is a direct result of the varying relative rate of gas depletion at
differing column densities.  For the lowest initial gas densities, the
evolutionary tracks follow an almost vertical trajectory as stars are formed at
very low efficiency.  In contrast, the higher initial gas densities form stars
very efficiently and thus the gas is quickly depleted relative to the star
formation time, $t_{SF}$.  We can easily rewrite Eqn.~\ref{sigrat} to describe
the star formation efficiency at $t=t_{SF}$ as:

\begin{equation}
\frac{\Sigma_*}{\Sigma_*+\Sigma_{gas}}(x,y,t_{SF}) = \frac{\Sigma_{gas}(x,y,t_{SF})}{\Sigma_{gas}(x,y,t_{SF})+1/(ckt_{SF} (1+\frac{t_{SF}}{t_0}))}
\end{equation}

\noindent At the $\Sigma_{gas}$ fiducial points of 20 and
300~$M_{\odot}$~pc$^{-2}$ adopted above,
$\frac{\Sigma_*}{\Sigma_*+\Sigma_{gas}}(x,y,t_{SF})~=~2.3\%$ and 26\%
respectively.  These are similar to reported values for ``distributed'' and
``clustered'' star formation \citep[e.g.][]{carp00,alle07,evan09}.  In this
analysis, as is claimed by \citet{bres10}, we see a smooth variation from low
to high column density star formation.  We add to that analysis that the most
likely progenitor of the smooth variation in stellar densities is the smooth
variation of gas column densities in star forming clouds.

\begin{figure}
\epsscale{.80}
\plotone{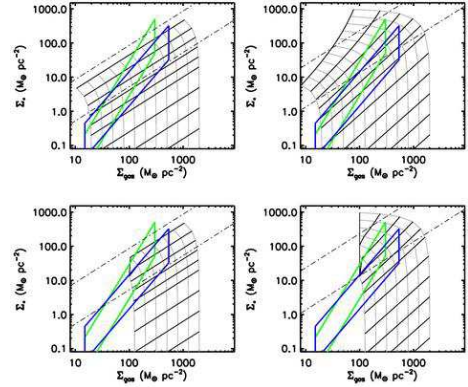}
\caption{Several other realizations of the model; the left column assumes an initial power law of slope 1 and $k = 0.1$~Myr$^{-1}$, and the right column assumes a slope of 1.4 and $k = 0.05$~Myr$^{-1}$~pc$^{0.8}~M_{\odot}^{-0.4}$; $c = 0.5$ in all realizations.  These values of $k$ were adopted to overlap with the observations, and do not affect the shape of the isochrones.  The black isochrone lines are adopted for the same time steps as used in Fig.~\ref{alpha2}.  The top row has no implicit gas column density threshold for star formation to occur.  The bottom row includes such a threshold at $100~M_{\odot}~$pc$^{-2}$; below that column density threshold, no stars are formed.  The MonR2 and Ophiuchus data extent polygons (green and blue, respectively) and $t=t_0$ and $t=0.1~t_0$ dot-dashed lines are overplotted as in Fig.~\ref{alpha2}.  The slope of 1 models are clearly inconsistent with the fits, and the slope of 1.4 with the star formation threshold is also inconsistent.  The slope of 1.4 model with no threshold is consistent with the Ophiuchus fit if some degree of evolution is applied, as the approximate effect is a steepening of the trend with time.\label{alpha}}
\end{figure}

In Fig.~\ref{alpha}, we investigate different values of $\alpha$, as well as
the possibility of a threshold in density for star formation, and compare the
results to the fits to the MonR2 and Ophiuchus clouds.  We show
Equation~\ref{fullsoln} models for $\alpha =1$ (top left panel) and $\alpha =
1.4$ (top right panel), with no threshold imposed, and with appropriate values
of $k$ ($\alpha$-dependent) to approximately fit the MonR2 and Ophiuchus data
(green and blue polygons, respectively): neither provides a good fit to the
data, unlike that in Fig~\ref{alpha2}.  We also implement a threshold (bottom
two panels) such that the star formation ceases when $\Sigma_{gas}(x,y,t)$ is
less than a value of 100~$M_{\odot}$~pc$^{-2}$.  A less extreme threshold has
been reported in a different analysis of the {\it Spitzer} c2d and Gould Belt
Legacy survey data, where protostars alone are used to compute star formation
surface densities \citep{heid10}.  Our analysis is not consistent with a star
formation threshold at any gas column density in the range that we have
surveyed.

\subsection{A Simple Model of Jeans Fragmentation\label{model}}

For MonR2 and Ophiuchus, the power law indices in the relation between stellar
and gas surface density are close to 2.  This section shows that such an
exponent is expected from a simple model of Jeans fragmentation in an
isothermal self-gravitating layer. 

If one initial Jeans mass makes one star, then the number surface density of stars in a layer equals the number surface density of Jeans masses,

\begin{equation}
			     N_* =  N_J
\label{jeans1}
\end{equation}

\noindent
The Jeans mass in a self-gravitating layer is

\begin{equation}
			M_J = \frac{B \sigma^4}{\Sigma_{gas}~G^2}
\label{jeans2}
\end{equation}

\noindent
where $B$ is a constant \citep[4.67 for an isothermal, self-gravitating layer of gas;][]{lars85}, $\sigma$ is the gas velocity dispersion, $\Sigma_{gas}$ is the gas mass surface density, and $G$ is the gravitational constant.

\noindent
Then the number surface density of Jeans masses can also be written as

\begin{equation}
			N_J  = \frac{\Sigma_{gas}}{M_J}
\end{equation}

\noindent
or substituting from \ref{jeans2},

\begin{equation}
			N_J = \frac{1}{B} (\frac{G \Sigma_{gas}}{\sigma^2})^2 
\end{equation}

\noindent
and substituting from \ref{jeans1}, multiplying both sides of the equation by a mean mass per star of 0.5~$M_{\odot}$\footnote{Note that the eventual mass of a given star is explicitly assumed to be uncorrelated with the local Jeans mass.  This is consistent with some characterizations of the pre-stellar core to stellar mass conversion process that include potential accretion of gas from the larger-scale environmental distribution \citep[e.g.][]{myer08,myer09b}.}, and converting to mass surface density units of $M_{\odot}$~pc$^{-2}$, with an assumed $\sigma = 0.2~$km~s$^{-1}$, the sound speed at 10~K:

\begin{equation}
			\Sigma_* = 1.34 \times 10^{-3} \Sigma_{gas}^2 
\end{equation}

Thus the power law of index 2 is predicted by simple thermal fragmentation of
an isothermal self-gravitating layer.  Furthermore, the value of $1.34 \times
10^{-3}$ is in the middle (in log-space) of our observed range of
$\Sigma_*/\Sigma_{gas}^2$ from section~\ref{corr}, and represents a mean value
for $ckt(1+t_{SF}/t_0)$ in the evolutionary model above.  We note that there is
no explicit time dependence in this model, thus although it gives the correct
star-gas correlation, it does not predict a rate of star formation.  

More generally, one must consider cloud layers that are non-uniform, spanning a
wide range of column densities in order to replicate the trends we report here,
which are found within each star-forming cloud individually.  An example of a
horizontally nonuniform planar system is the density-modulated isothermal layer
\citep{schm67}, or of a planar system of isothermal cylinders \citep{ostr64}
can be expected to give a similar relation between stellar and gas surface
densities.  Some of the clouds studied in this paper may have such a flattened
geometry; their extinction maps indicate that ``hub-filament'' structure is
expected if a clumpy medium is compressed into a modulated self-gravitating
layer \citep{myer09}.

If this picture of fragmentation in a layer applies to some of the clouds
considered here, the results of Section~\ref{tommodel} indicate that
evolutionary effects probably do not change the relation between star and gas
surface density very much, over the likely star-forming life of the cloud.

\subsection{Empirical Comparison to Cloud-Scale Values, the Kennicutt-Schmidt Law, and Recent Parsec-scale Work}

\begin{figure}
\plotone{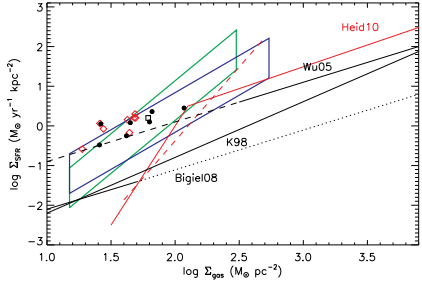}
\caption{Our reproduction of the Star Formation Rate Surface Density vs. Gas Column Density figure from \citet{evan09}.  The filled black circles are mean measurements for entire molecular clouds from the c2d survey as presented in that work, and the open square is their mean.  Red diamonds are similar cloud-integrated values derived from the data compiled in Table~\ref{littab1}.  The black lines are some recent correlations reported in the literature \citep{kenn98,wu05,bigi08} based largely on extra-galactic measures (500~pc or larger size scales).  Dashed portions of the line are extrapolated relative to the reported range of each trend.  The green and blue polygons demonstrating the correlations reported for MonR2 and Ophiuchus, respectively, have been converted to star formation rate density under identical assumptions to \citet{evan09} and overlaid.  The cloud integrated values are consistent with our measurements, but the latter extend to higher column density because of the smaller size scales considered.  The red trends are from \citet{heid10}; the solid line is their bent power law fit to a combined set of {\it Spitzer}- and {\it IRAS}--derived data, and the dashed line is our fit to their {\it Spitzer}-derived data only.\label{neal1}}
\end{figure}

As already noted, \citet{evan09} compared the {\it overall} star formation
surface density vs. gas column density for each of the clouds in the c2d
survey, as well as an average of all of them, for comparison with the
Kennicutt-Schmidt law for galaxies.  They find that the star formation rate
density predicted by the law falls a factor of 20 below their observed data.
In Figure~\ref{neal1}, we have overlaid the correlation reported here for even
smaller scales (0.3 to 2.0~pc) on our reproduction of the \citet{evan09}
figure.  Our correlation overlays the \citet{evan09} cloud-averaged data well
at intermediate surface densities, and the mean value for all clouds falls
quite close to our correlation.  The difference is only clear where our trend
extends to higher column densities.  Because we are able to probe a wider range
in gas column density by taking measurements at smaller size scales, it is not
surprising that such a discrepancy could be uncovered.  Regardless, our
analysis agrees with that of \citet{evan09}; studies of star formation rate
density versus gas density for nearby clouds show much greater star formation
rate for the local gas density than extragalactic correlations would suggest.
As noted by \citet{evan09} and others, the measurements used to derive most
star formation law characterizations vary widely in the size scales and time
scales probed, and these variations almost certainly play a role in the
apparent discrepancies.  Indeed, recent work to characterize the role of
measurement size scale in characterizing the correlation between star formation
rate surface densities and gas surface densities in nearby galaxies has
revealed considerable change in the power law index with the measurement size
scale used \citep{liu11}.  

Other recent efforts have examined the star formation rate and gas surface
density correlation in nearby star-formation regions via similar sorts of data
(ie. YSO star counts and near-IR extinction maps) and some overlap in the
specific clouds considered.  
\citet{lada10} estimated the star formation rate vs. cloud mass for a sample of
11 nearby clouds.  Within this cloud sample, they noted that there was a wide
dispersion in the estimated star formation rate per cloud mass.  They then
demonstrated that the observed dispersion  within their cloud sample was
minimized if they only considered the cloud mass at column densities above $A_V
= 7$ (or 116~$M_{\odot}$~pc$^{-2}$).  They proposed that this corresponded to a
density threshold for star formation; for densities above this threshold the
star formation rate varied linearly with gas mass, below the threshold the star
formation rate was negligible.

A similar threshold ($129 \pm 14~M_{\odot}$~pc$^{-2}$) was invoked by the
sub-cloud-scale analysis by \citet{heid10}, a follow-up to the cloud-scale
examination by \citet{evan09} mentioned above.  \citet{heid10} used the numbers
of {\it Spitzer}-identified Class~I and so-called ``flat spectrum'' YSOs (ie.
they excluded the longer-lived Class~II YSOs) within various extinction contour
intervals to probe the correlation in 20 nearby molecular clouds.  
Since their {\it Spitzer} data did not include clouds forming massive stars, they did not have a direct measurement of YSO densities in high mass star
forming regions.
Thus they
added star formation rates derived from {\it IRAS} far-IR luminosities and mean
gas column densities derived from HCN line emission for a set of more distant
and presumably more active star-forming regions.  We have overlaid their final
bent power law best fit to their combined data in Fig.~\ref{neal1} (red solid
lines), showing that it underestimates the star formation rates we report here
at both extremely high and low gas column densities.  

At high column densities, the shallow portion of the bent power law of
\citet{heid10} is driven by the inclusion of the {\it IRAS}-HCN-derived data in
the fit; their {\it Spitzer}-derived data seem relatively consistent with our
analysis (examined in more detail below).  That work includes a discussion of
the likelihood that far-IR-derived star formation rates may be underestimated
for individual star-forming regions by up to an order of magnitude.  We find
plenty of evidence for that to be the case in the nearby embedded clusters,
where we can do direct comparisons between the {\it IRAS} far-IR-derived star
formation rate and that derived from the YSO counts.  For example, using the
known far-IR luminosity \citep[26,000~$L_{\odot}$;][]{ridg03} and the YSO count
for the cluster core of the MonR2 main cloud core \citep[132 YSOs, 0.73~pc
effective radius;][]{gute09}, we obtain star formation rate surface densities
of 3 and 20~$M_{\odot}$~yr$^{-1}$~kpc$^{-2}$, respectively.  A similar
discrepancy can be found in the Orion Nebula Cluster.  Its far-IR luminosity is
estimated at $10^5~L_{\odot}$ \citep{geza98}, yet its central parsec contains
of order $10^3$ YSOs \citep[e.g.][]{feig05}, yielding SFR surface density
estimates of 20 and $300~M_{\odot}$~yr$^{-1}$~kpc$^{-2}$, respectively.  In
summary, the common conversion of far-IR flux to star formation rate appears to
underestimate the true star-formation rate at parsec scales in nearby regions.



We are not concerned with the apparent discrepancy with our result at low gas
column densities. The protostellar number counts \citep{heid10} at these gas
column densities are very low: a star surface density of
$10^{-1.7}~M_\odot$~kpc$^{-2}$~yr$^{-1}$ corresponds to one $0.5~M_\odot$
protostar per 0.5~Myr (the estimated lifetime of the evolutionary phase) in the
area of an entire typical molecular cloud ($\sim10$~pc size scale is typical of
nearby clouds).  
We used the fitting technique described in Section~\ref{corr} to fit their {\it
Spitzer}-derived data, excluding the upper limits.  We find that the {\it
Spitzer}-only correlation to be a power law of slope $3.7 \pm 1.6$ (dashed red
line in Fig.~\ref{neal1}), a steeper power law than our reported trend (though
within $1\sigma$) and lacking in an apparent gas surface density threshold for
star formation.  Differences at this level may be explained in part by
differences in source classification method and uncertainty in the relative
durations of the pertinent YSO evolutionary stages, but detailed consideration
of such effects is beyond the scope of this paper.

\section{Summary\label{summary}}

We have presented an analysis of YSO and gas surface density at 0.3 to 2.0 pc
scales for over 7000 YSOs in eight nearby molecular clouds from 40 square
degrees of {\it Spitzer}- and 2MASS--surveyed sky.  These plots show the
relationship between the surface density of stars formed over the lifetime of
the molecular cloud vs. the surface density of the remaining gas in the clouds.
A power law correlation between these two quantities is evident in all eight
molecular clouds included in the study.  In summary:

\begin{itemize}

\item MonR2 and Ophiuchus have particularly strong and well-defined
correlations, thus we fit them each with power laws (indexes of 2.67 and 1.87,
respectively).

\item Some clouds exhibit large deviations from the power law form apparent in
MonR2 and Ophiuchus; we argue that these deviations are due to gas dispersal by
massive stars and non-coevality in each cloud.  The
presence of non-coeval regions in the clouds is demonstrated by
variations in the ratio of the numbers of Class~II to Class~I YSOs.

\item Even given non-coevality and gas dispersal in most clouds, we
find that greater than half of all YSOs with excess infrared emission in every
cloud considered are found to have local $\Sigma_*/\Sigma^2_{gas}$ in the range
of $3 \times 10^{-4}$ and $5 \times 10^{-3}$~pc$^2$~M$^{-1}_{\odot}$.

\item Extracting the regions of each cloud found within a common evolutionary
state ( $3 \times 10^{-4} < \Sigma_*/\Sigma^2_{gas} < 5 \times 10^{-3}$ ), the
Class~II to Class~I ratio is shown to decline with gas column density.
However, the magnitude of the decline varies by cloud, from near-uniform ratios
with small declines in MonR2 and North America Nebula to steep declines in
S140 and Perseus.

\item  The correlation can be reproduced by a simple model where the star
formation rate per unit area varies with the square of the surface density of
gas.  This model accounts for the gas removed from the cloud by the accretion
and ejection of gas mass by each star.  We find that the accretion and ejection
of gas will cause deviation from the power law form, but will only result in a
mild steepening of the correlation at high gas surface densities over the
likely star formation timescales of a few Myr.  In contrast, we find star
formation laws where the star formation rate per area varies linearly or with
the 1.4 power of the gas surface density are not able to fit the observed
correlation.

\item The observed dependence of the stellar surface density with the gas
density squared is a prediction of the thermal fragmentation of a sheet-like
isothermal layer.  To achieve the observed range of local column densities, one
must invoke some sort of nonuniformity in the local column density structure of
the sheet, such as a ``hub-filament'' geometry as described and analytically
characterized by \citet{myer09}.  

\item The correlation reported here is largely consistent with recent {\it
Spitzer}-derived estimates of star formation rate and gas surface densities
both averaged over clouds \citep{evan09} and measured within them
\citep{heid10}.  We have identified some systematic effects that may explain
differences between our results and the analysis of \citet{heid10} that probes
comparable size scales.  Regardless of these differences, all of the work
considered consistently suggests that extragalactic star formation rate and gas
surface density correlations significantly underestimate those seen in nearby
clouds.  

\end{itemize}

\appendix

\section{Expansion of the c2d Extinction Maps\label{app1}}

The c2d extinction maps are generated using extinction measurements from each
source classified as having photospheric colors in their merged 2MASS and {\it
Spitzer} catalogs.  The measurement is a parameter returned as part of a
blackbody fit to the 1-8$\mu$m spectral energy distributions of those sources,
limited by what detections where achieved to reasonable S/N in that range.
Unfortunately, that means that the c2d extinction maps are limited to the field
of view at least covered by IRAC.  Since the c2d survey areas were chosen to
match elevated extinction in the surveyed clouds ($A_V>3$), obtaining an
unbiased local baseline is impossible from the maps directly.  This baseline is
needed to remove the foreground extinction to the reddening map.

In order to obtain a reasonable baseline $A_V$ offset for the clouds surveyed
by c2d, we had to expand those maps such that they extend beyond the
IRAC-surveyed region, which was largely constrained to $A_V>3$ areas in each
cloud.  Our process is as follows:

\begin{itemize}

\item Construct an $A_V$ map of a large area around each cloud using the technique from \citet{gute09} that was used for the non-c2d clouds included in this work.

\item Resample the large map onto the c2d map's grid. 

\item Fit the $A_V(c2d)$ vs. $A_V(G09)$ correlation between the two maps using those pixels that have valid data in both maps, in the range where both maps should have strong agreement: $2<A_V(c2d)<10$.

\item Convert the large map's $A_V$ values to use the c2d map's effective reddening law and zero-point calibration using the fitted parameters and fill in the empty parts of the c2d map's grid.

\end{itemize}

\begin{figure}
\plotone{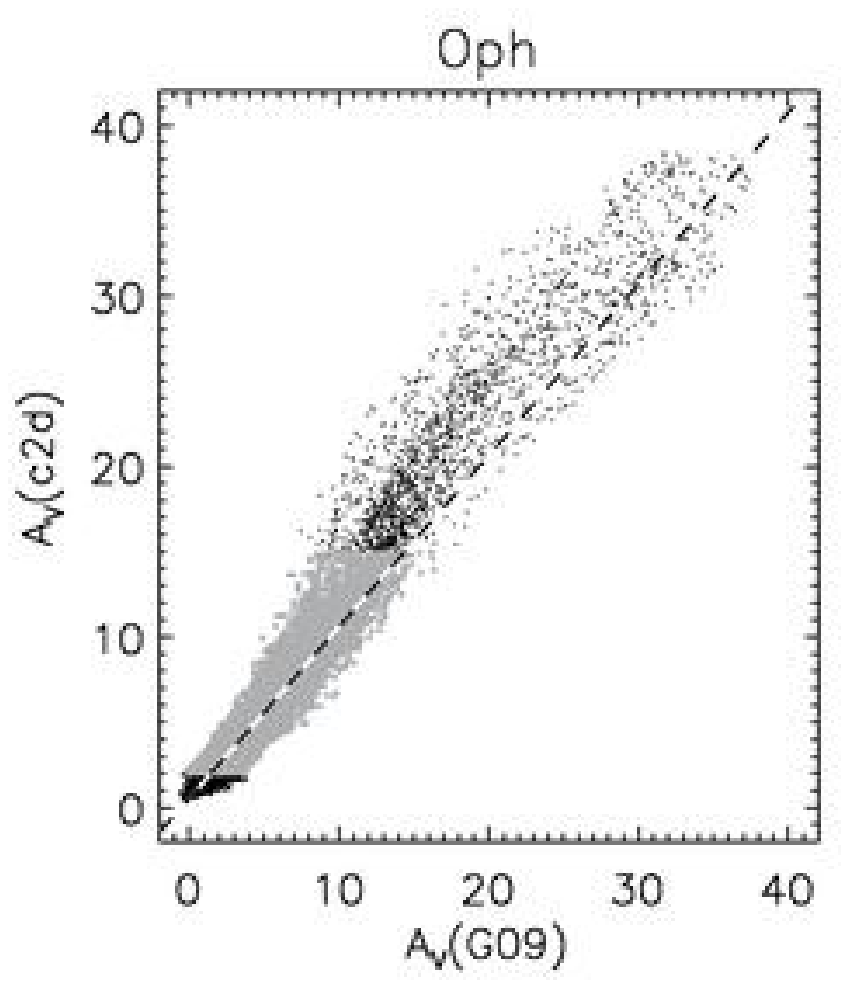}
\caption{$A_V$ comparison between the c2d-generated map of Ophiuchus and our own, resampled to their grid.  The fit was performed on the gray points only, where we expect both maps to have strong agreement.\label{c2dcomp_oph}}
\end{figure}

\begin{figure}
\plotone{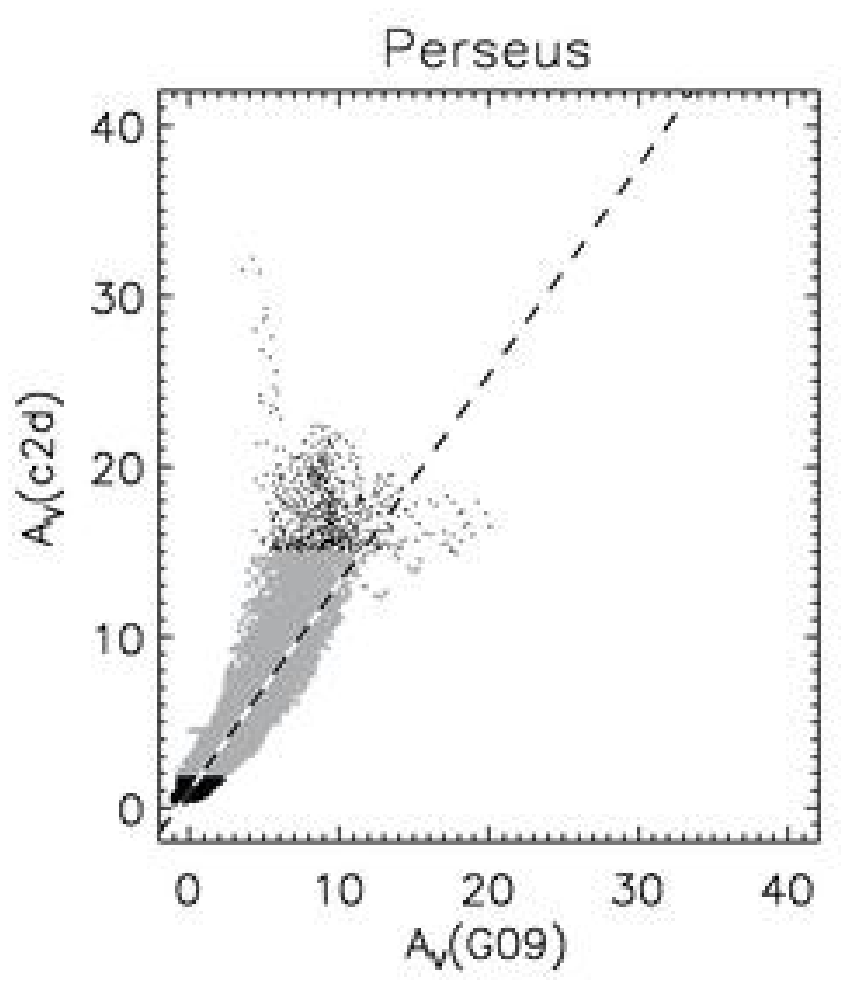}
\caption{$A_V$ comparison between the c2d-generated map of Perseus and our own, resampled to their grid.  The fit was performed on the gray points only, where we expect both maps to have strong agreement.\label{c2dcomp_per}}
\end{figure}

\begin{figure}
\plotone{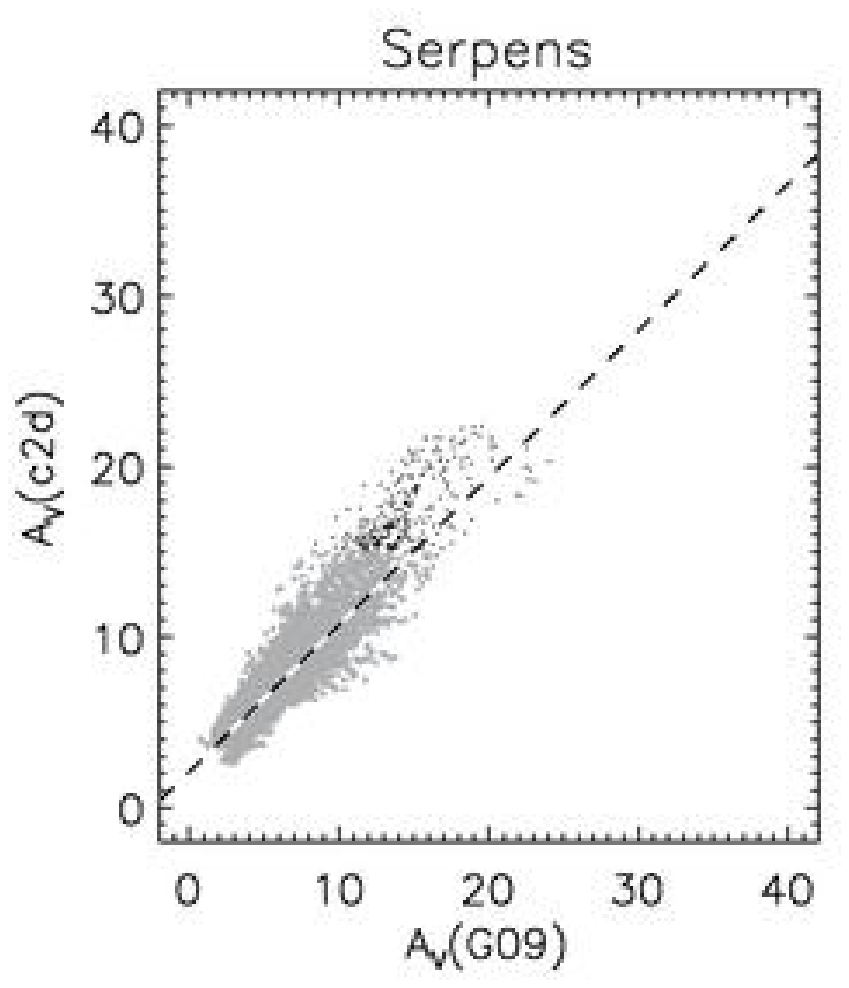}
\caption{$A_V$ comparison between the c2d-generated map of Serpens and our own, resampled to their grid.  The fit was performed on the gray points only, where we expect both maps to have strong agreement.\label{c2dcomp_serp}}
\end{figure}

The fits are shown in
Figs.~\ref{c2dcomp_oph},~\ref{c2dcomp_per},~\&~\ref{c2dcomp_serp}, and the
results are summarized in Table~\ref{c2dcomp}.  The figures reveal a consistent
deviation and spreading in the data for $A_V(c2d) > 12$ where the $A_V(c2d)$
values are generally higher than the fit.  The declining background star
density with increasing extinction forces the adaptive smoothing of the G09
method to smooth over poorly sampled high column density structure, diluting it
in favor of higher statistical S/N.  In addition, the inclusion of the
IRAC-derived data in the c2d maps likely enables higher S/N extinction
measurements in greater numbers in intermediate extinction regions.  

The fit results are summarized in Table~\ref{c2dcomp}.  The inclusion of IRAC
data in the c2d extinction measurement process could lead to deviations in the
effective reddening law; indeed, color excess ratios in various combinations of
IRAC and 2MASS bandpasses have been shown to vary slightly \citep[e.g. Serpens
and Ophiuchus;][]{flah07}.  That said, these effects are expected to be small,
and that is born out by the $<15\%$ variation indicated by the near-unity
slopes of the fits.  The offset calibration deviations are easily explained by
the use of local reference fields in the c2d maps to establish the typical
colors of field stars near each cloud.  In constrast, the G09 technique adopts
an intrinsic $H-K=0.2$ color for field stars and then utilizes constant
baseline subtraction away from the cloud to account for local variations.  With
our cross calibration, we ensure a fair baseline treatment for the c2d clouds,
regardless of the differences in their construction.  

Additionally, we can test how the extinction mapping technique affects the
log~$\Sigma_*$~vs.~log~$\Sigma_{gas}$ fit to the Ophiuchus cloud; the power law
index is $1.93 \pm 0.03$ with reduced $\chi^2$ of 5.01 using the map generated
by our technique, and $1.87 \pm 0.03$ with reduced $\chi^2$ of 4.78 using the
c2d map.  Thus the values differ by less than 10\% and are within 1.5 sigma of each other.

\acknowledgments

This publication makes use of data products from the Two Micron All Sky Survey,
which is a joint project of the University of Massachusetts and the Infrared
Processing and Analysis Center/California Institute of Technology, funded by
the National Aeronautics and Space Administration and the National Science
Foundation.  This research has made use of the SIMBAD database, operated at
CDS, Strasbourg, France.  This research has made use of the VizieR catalogue
access tool, CDS, Strasbourg, France.  This work is based in part on
observations made with the Spitzer Space Telescope, which is operated by the
Jet Propulsion Laboratory, California Institute of Technology under a contract
1407 with NASA.  Support for the IRAC instrument was provided by NASA through
contract 960541 issued by JPL.

{\it Facilities:} \facility{Spitzer}.

\begin{deluxetable}{cccccccccccc}
\rotate
\tabletypesize{\scriptsize}
\tablecaption{Data and Derived Properties by Cloud\label{littab1}}
\tablewidth{0pt}
\tablehead{\colhead{Name} & \colhead{Mass\tablenotemark{a} $(M_{\odot})$} & \colhead{Area (pc$^2$)} & \colhead{Resolution (pc)} & \colhead{Baseline $A_V$} & \colhead{Mean $A_V$\tablenotemark{a}} & \colhead{Max. $A_V$\tablenotemark{a}} & \colhead{$N_{II}$} & \colhead{$N_I$} & \colhead{Dist. (pc)} & \colhead{Dist. Ref.} &\colhead{YSO Ref.}}
\startdata
MonR2 &  25800 &  1170.0 & 0.95 & -0.5 &  1.5 & 17.1 & 815 & 188 &  830 & 1 & 9 \\
CepOB3 &  16100 &   551.0 & 0.42 &  0.9 &  1.9 & 21.6 & 1976 & 196 &  700 & 2 & 9 \\
S140 &   7150 &   200.0 & 0.51 &  0.8 &  2.4 & 21.3 & 470 & 56 &  750 & 3 & 9 \\
Orion &  33200 &   572.0 & 0.75 & -1.0 &  3.9 & 24.1 & 2861 & 502 &  414 & 4 & 10 \\
NANeb &  24400 &   548.0 & 0.29 &  1.1 &  3.0 & 22.0 & 745 & 425 &  600 & 5 & 5 \\
Oph &   3070 &    43.0 & 0.20 &  0.2 &  4.8 & 38.0 & 179 & 71 &  150 & 6 & 11\tablenotemark{b} \\
Perseus &   4300 &    72.8 & 0.29 &  0.7 &  3.9 & 31.5 & 243 & 111 &  250 & 7 & 11\tablenotemark{b} \\
Serpens &   2590 &    45.0 & 0.48 &  3.7 &  3.8 & 18.7 & 141 & 54 &  415 & 8 & 11\tablenotemark{b} \\
\enddata
\tablerefs{(1) \citet{carp00}; (2) \citet{mosc09}; (3) \citet{hiro08}; (4) \citet{ment07}; (5) \citet{rebu11}; (6) \citet{wilk05}; (7) \citet{enoc06}; (8) \citet{dzib10}; (9) \citet{gute11}; (10) \citet{mege11}; (11) \citet{evan09} }
\tablenotetext{a}{Integrated mass and mean and maximum $A_V$ values are derived from the baseline-subtracted extinction maps for the areas within the {\it Spitzer} coverage.}
\tablenotetext{b}{Minor source count discrepancies between our reported counts and Table 5 of \citet{evan09} have been confirmed by the authors as a mistake in their table.  Under the their advisement, we have included counts derived from the source catalogs of \citet{evan09}, and their Table 5 source counts summary will be corrected in a forthcoming paper on all of the Gould Belt clouds (M. Dunham, private communication).}
\end{deluxetable}

\begin{deluxetable}{ccccc}
\tabletypesize{\scriptsize}
\tablecaption{log~$\Sigma_*$ vs. log~$\Sigma_{gas}$ Fit Coefficients\label{resulttab0}}
\tablewidth{0pt}
\tablehead{\colhead{Name} & \colhead{Pearson Index} & \colhead{Calibration Offset} & \colhead{Power Law Index} & \colhead{Reduced $\chi^2$}}
\startdata
MonR2 & 0.87 &  -4.42 & $2.67 \pm 0.02$ & 6.22 \\
CepOB3 & 0.17 & -2.21 & $1.77 \pm 0.01$ & 20.3 \\
S140 & 0.52 & -1.84 & $1.37 \pm 0.03 $ & 7.81 \\
Orion & 0.60 & -2.40 & $1.80 \pm 0.01$ & 17.9 \\
NANeb & 0.67 & -3.54 & $2.15 \pm 0.02$ & 12.2 \\
Ophiuchus & 0.83 & -3.10 & $1.87 \pm 0.03$ & 4.78 \\
Perseus & 0.59 & -7.0 & $3.8 \pm 0.1$ & 13.0 \\ 
Serpens & 0.83 & -3.05 & 1.95 & \tablenotemark{a} \\
\enddata
\tablenotetext{a}{The Serpens fit's $\chi^2$ minimization did not converge using the \citet{flah07} fitting technique; a simpler, unweighted bisector fit has been used to obtain the reported coefficients.  Saturation in the gas column density appears to cause a dual valued signal in the high column density region of the trend.}
\end{deluxetable}

\begin{deluxetable}{ccc}
\tabletypesize{\scriptsize}
\tablecaption{$\Sigma_* / \Sigma^2_{gas}$ Median Values by YSO Class\label{resulttab1}}
\tablewidth{0pt}
\tablehead{\colhead{Name} & \colhead{Class~II} & \colhead{Class~I}}
\startdata
MonR2 &  $0.86 \times 10^{-3}$& $0.68 \times 10^{-3}$\\
CepOB3 &  $3.0 \times 10^{-3}$&  $1.0 \times 10^{-3}$ \\
S140 &   $1.7 \times 10^{-3}$  & $1.2 \times 10^{-3}$\\
Orion &  $1.7 \times 10^{-3}$&  $1.0 \times 10^{-3}$\\
NANeb &   $0.64 \times 10^{-3}$ &  $0.81 \times 10^{-3}$\\
Ophiuchus &  $ 0.47 \times 10^{-3}$& $0.40 \times 10^{-3}$\\
Perseus &  $1.9 \times 10^{-3}$ & $0.90 \times 10^{-3}$\\
Serpens &  $0.76 \times 10^{-3}$ & $0.57 \times 10^{-3}$  \\
\enddata
\end{deluxetable}

\begin{deluxetable}{ccc}
\tabletypesize{\scriptsize}
\tablecaption{Extinction Comparison Fit Coefficients\label{c2dcomp}}
\tablewidth{0pt}
\tablehead{\colhead{Name} & \colhead{Constant Offset ($A_V$)} & \colhead{Slope}}
\startdata
Oph &  0.666 &  $0.959 \pm 0.001$ \\
Perseus &  1.127 & $1.126 \pm 0.004$  \\
Serpens &  1.633 & $0.922 \pm 0.005$  \\
\enddata
\end{deluxetable}

\begin{figure}
\epsscale{.80}
\plotone{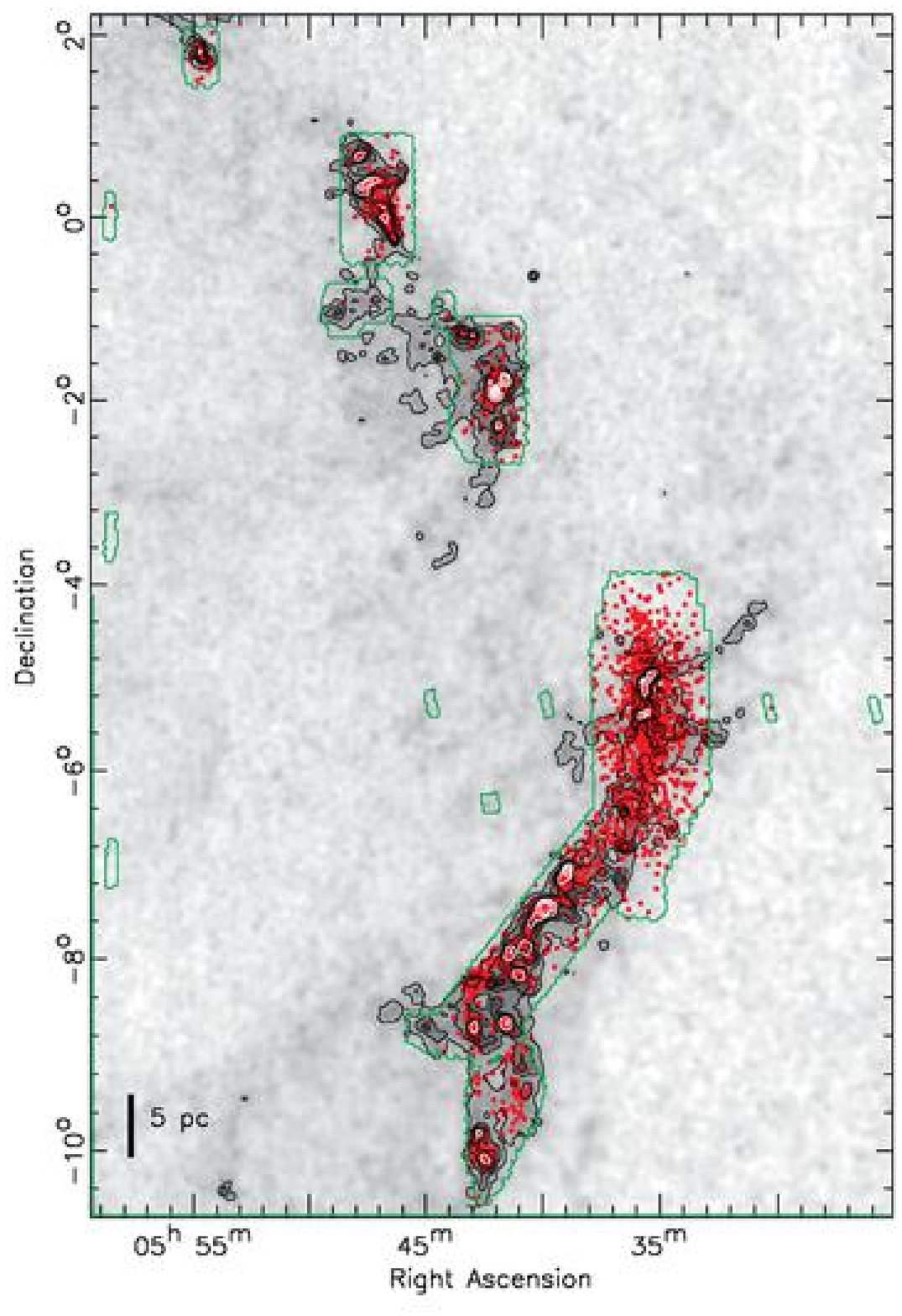}
\caption{Extinction map of the Orion cloud overlaid in red with the spatial distribution of {\it Spitzer}-identified YSOs.  The grayscale and contour properties are identical to those in Fig.~\ref{monr2_im}.}
\end{figure}

\begin{figure}
\epsscale{.80}
\plotone{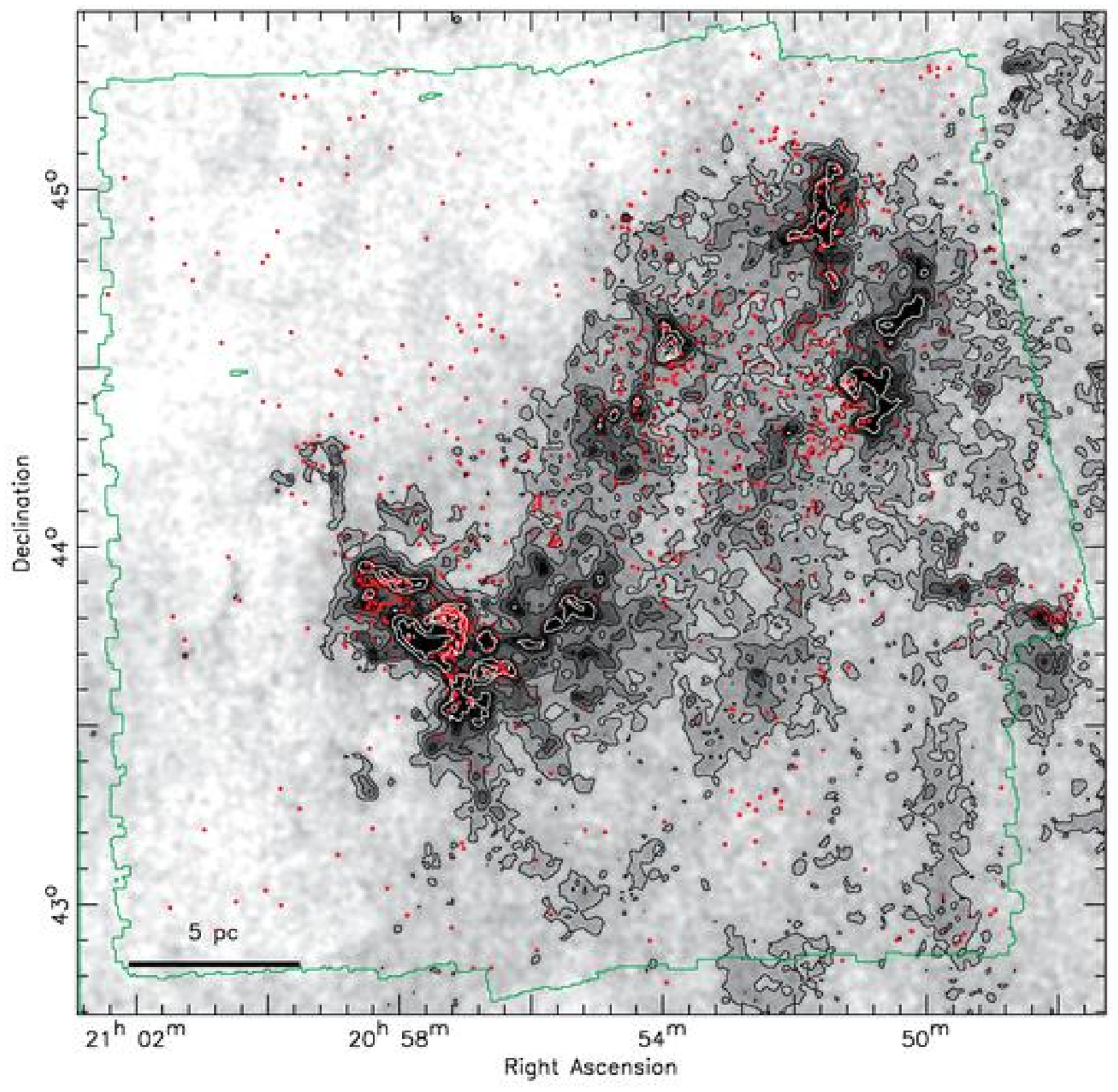}
\caption{Extinction map of the NANeb cloud overlaid in red with the spatial distribution of {\it Spitzer}-identified YSOs.  The grayscale and contour properties are identical to those in Fig.~\ref{monr2_im}.}
\end{figure}

\begin{figure}
\epsscale{.80}
\plotone{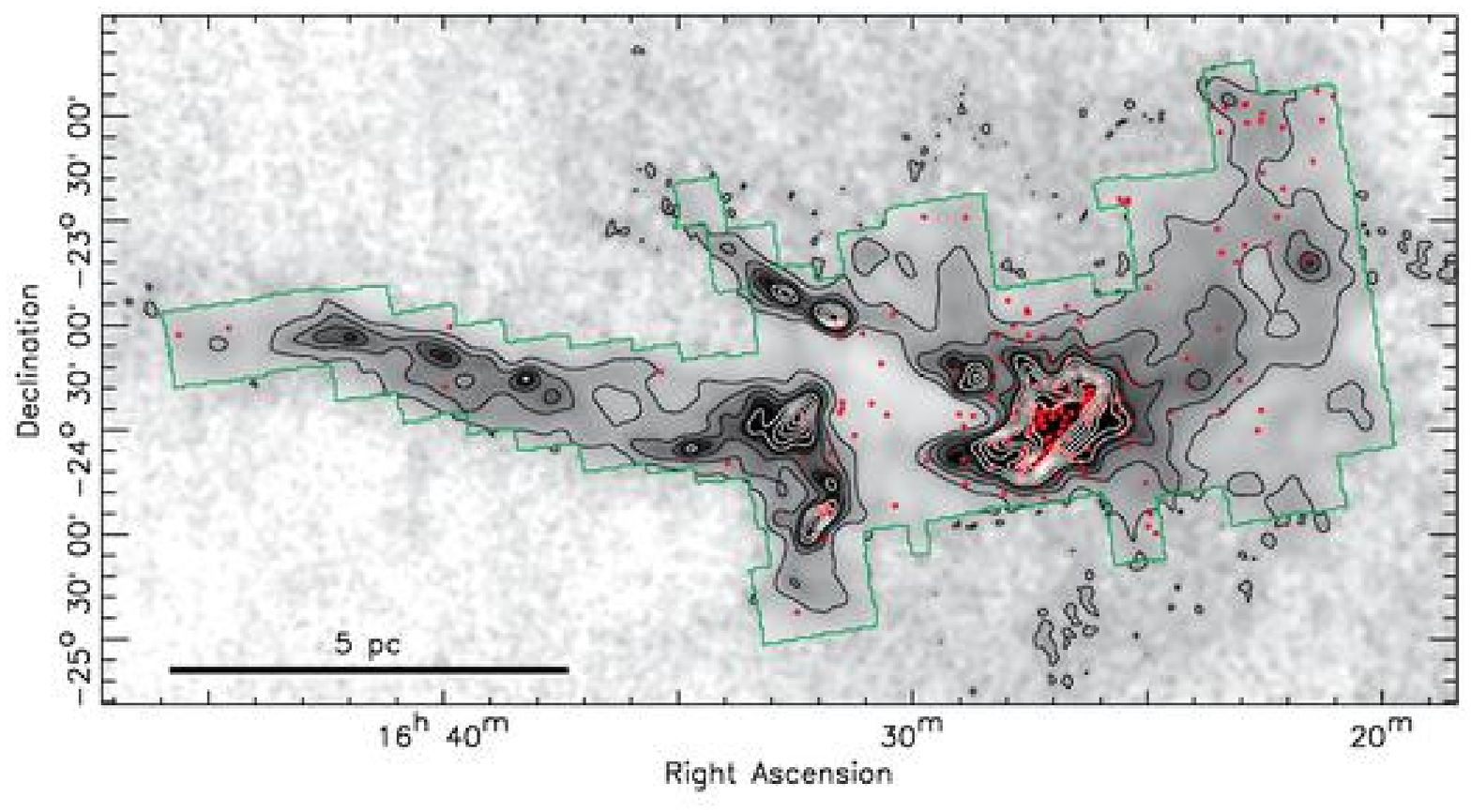}
\caption{Extinction map of the Ophiuchus cloud overlaid in red with the spatial distribution of {\it Spitzer}-identified YSOs.  The grayscale and contour properties are identical to those in Fig.~\ref{monr2_im}.}
\end{figure}

\begin{figure}
\epsscale{.80}
\plotone{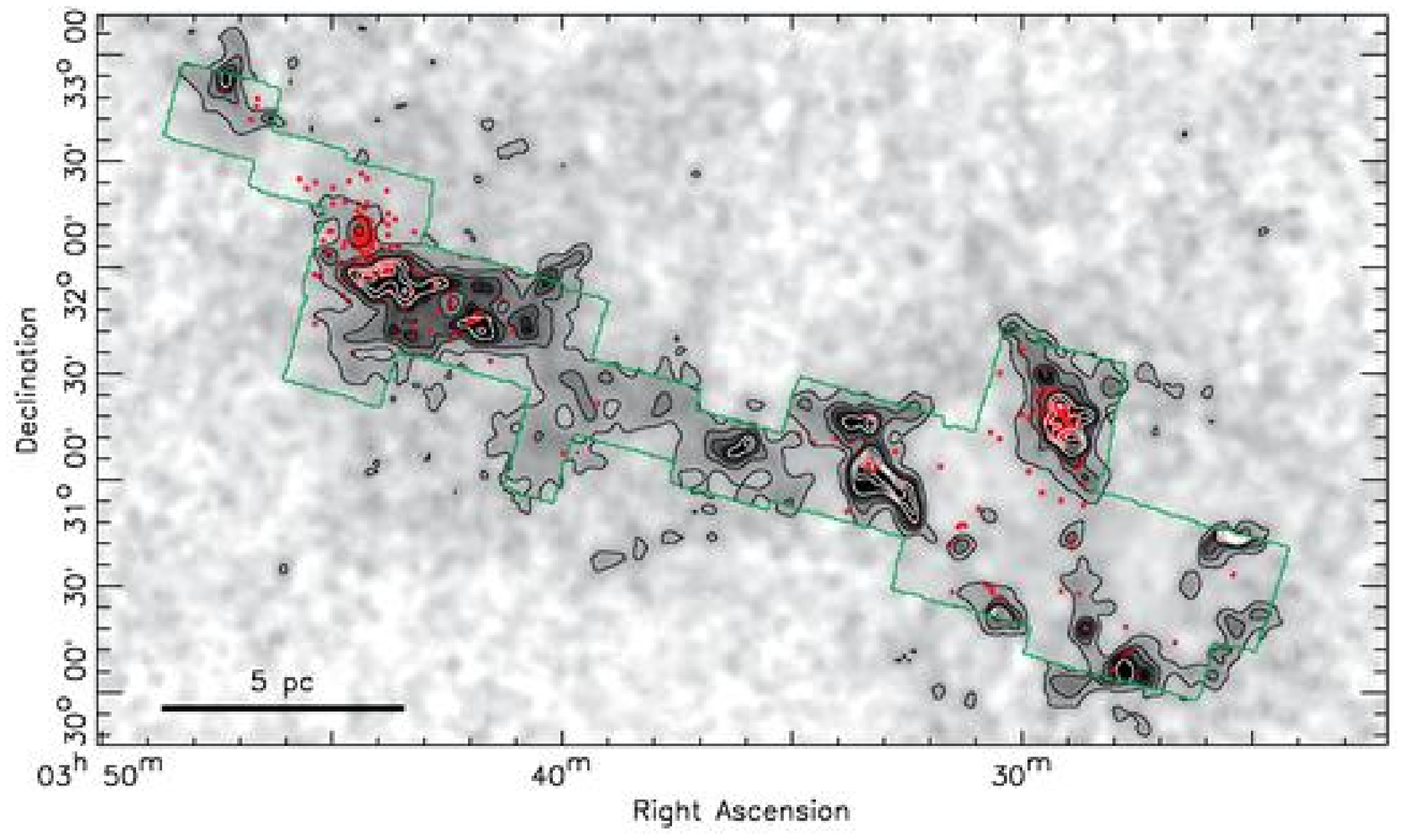}
\caption{Extinction map of the Perseus cloud overlaid in red with the spatial distribution of {\it Spitzer}-identified YSOs.  The grayscale and contour properties are identical to those in Fig.~\ref{monr2_im}.}
\end{figure}

\end{document}